\documentclass[useAMS,usenatbib]{mn2e} 

\usepackage{amsmath}
\usepackage{amssymb}
\usepackage{amsfonts}
\usepackage{graphicx}
\usepackage{txfonts}
\usepackage{natbib}
\usepackage{longtable}
\usepackage{multirow}

\newcommand{\fluxunits}{{\rm erg}\;{\rm s}^{-1}{\rm cm}^{-2}}
\newcommand{\kbol}{k_{\rm bol}}
\newcommand{\Lbol}{L_{\rm bol}}
\newcommand{\Lir}{L_{\rm IR}}
\newcommand{\Lx}{L_{\rm X}}
\newcommand{\Lhard}{L_{[2-10]{\rm keV}}}
\newcommand{\Lsoft}{L_{[0.5-2]{\rm keV}}}

\newcommand{\Log}{{\rm Log}~}
\newcommand{\NH}{N_{\rm H}}
\newcommand{\lambdaEdd}{\lambda_{\rm Edd}}
\newcommand{\Mbh}{M_{\rm BH}}

\newcommand{\Ledd}{L_{\rm Edd}}
\newcommand{\alphaox}{\alpha_{\rm ox}}
\newcommand{\cpdf}{p(\Lx\mid M_\ast,z)}
\newcommand{\rev}[1]{{ #1}}

\newcommand{\lo}{\rm L_{2500 \;\text{\AA}}}
\newcommand{\lx}{\rm L_{2 \;{\rm keV}}}

\DeclareRobustCommand{\ion}[2]{%
\relax\ifmmode
\ifx\testbx\f@series
{\mathbf{#1\,\mathsc{#2}}}\else
{\mathrm{#1\,\mathsc{#2}}}\fi
\else\textup{#1\,{\mdseries\textsc{#2}}}%
\fi}

\title[Bolometric luminosities and Eddington ratios of AGN in XMM-COSMOS]{Bolometric luminosities and Eddington ratios of X--ray selected Active Galactic Nuclei in the XMM-COSMOS survey}
\author[E. Lusso et al.]{E.~Lusso$^{1,2}$, A.~Comastri$^{2}$, B.~D.~Simmons$^{3,4}$, M.~Mignoli$^{2}$, G.~Zamorani$^{2}$, C.~Vignali$^{2,5}$, 
\newauthor M.~Brusa$^{6}$, F.~Shankar$^{7}$, D. Lutz$^{6}$, J.~R.~Trump$^{8}$, R.~Maiolino$^{9,10}$, R.~Gilli$^{2}$, M.~Bolzonella$^{2}$, 
\newauthor S.~Puccetti$^{11}$, M.~Salvato$^{6,12}$, C.~D.~Impey$^{13}$, F.~Civano$^{14}$, M.~Elvis$^{14}$, V.~Mainieri$^{15}$, 
\newauthor J.~D.~Silverman$^{16}$, A. M. Koekemoer$^{17}$, A. Bongiorno$^{9}$, A. Merloni$^{6,12}$, S. Berta$^{6}$,  
\newauthor E. Le Floc'h$^{18}$, B. Magnelli$^{6}$, F. Pozzi$^{5}$ and L. Riguccini$^{18}$ \\
$^{1}$Max Planck Institut f\"{u}r Astronomie, K\"{o}nigstuhl 17, D-69117, Heidelberg, Germany. e-mail:~\textsl{lusso@mpia.de}\\
$^{2}$INAF--Osservatorio Astronomico di Bologna, via Ranzani 1, I-40127 Bologna, Italy.\\
$^{3}$Astronomy Department, Yale University, New Haven, CT 06511, USA.\\
$^{4}$Yale Center For Astronomy \& Astrophysics, Physics Department, Yale University, New Haven, CT 06511, USA.\\
$^{5}$Dipartimento di Astronomia, Universit\`{a} di Bologna, via Ranzani 1, I-40127 Bologna, Italy.\\
$^{6}$Max Planck Institut f\"{u}r Extraterrestische Physik, Postfach 1312, 85741 Garching bei M\"{u}nchen, Germany.\\
$^{7}$GEPI, Observatoire de Paris, CNRS, Univ. Paris Diderot, 5 Place Jules Janssen, F-92195 Meudon, France.\\
$^{8}$University of California Observatories/Lick Observatory, University of California, Santa Cruz, CA 95064.\\
$^{9}$INAF - Osservatorio Astronomico di Roma, via di Frascati 33, Monte Porzio Catone (RM), 00040, Italy.\\
$^{10}$Cavendish Laboratory, University of Cambridge, 19 J. J. Thomson Ave., Cambridge CB3 0HE,  UK.\\
$^{11}$ASI Science Data Center, via Galileo Galilei, 00044 Frascati, Italy.\\
$^{12}$Max Planck Institut f\"{u}r Plasma Physik and Excellence Cluster, 85748 Garching, Germany.\\
$^{13}$Steward Observatory, University of Arizona, 933 North Cherry Avenue, Tucson, AZ 85721, USA.\\
$^{14}$Harvard-Smithsonian Center for Astrophysics, 60 Garden Street,Cambridge, MA 02138, USA.\\
$^{15}$ESO, Karl-Schwarzschild-Strasse 2, 85748 Garching bei M\"{u}nchen, Germany.\\
$^{16}$Institute for the Physics and Mathematics of the Universe (IPMU), University of Tokyo, Kashiwanoha 5-1-5, Kashiwashi, Chiba 277-8568, Japan.\\
$^{17}$Space Telescope Science Institute, Baltimore, Maryland 21218, USA.\\
$^{18}$Laboratoire AIM, CEA/DSM-CNRS-Universit\'{e} Paris Diderot, IRFU/Service d'Astrophysique, B\^{a}t.709, CEA-Saclay, 91191 Gif-sur-Yvette Cedex, France.\\
}
\date{Accepted version June 12, 2012}

\begin{document}
   \maketitle
%
\begin{abstract}
Bolometric luminosities and Eddington ratios of both X--ray selected broad-line (Type-1) and narrow-line (Type-2) AGN from the XMM-\textit{Newton} survey in the COSMOS field are presented. 
The sample is composed by 929 AGN (382 Type-1 AGN and 547 Type-2 AGN) and it covers a wide range of redshifts, X--ray luminosities and absorbing column densities. About 65\% of the sources are spectroscopically identified as either Type-1 or Type-2 AGN (83\% and 52\% respectively), while accurate photometric redshifts are available for the rest of the sample.
The study of such a large sample of X-ray selected AGN with a high quality multi-wavelength coverage from the far-infrared (now with the inclusion of \textit{Herschel} data at 100~$\mu$m and 160~$\mu$m) to the optical-UV allows us to obtain accurate estimates of bolometric luminosities, bolometric corrections and Eddington ratios. 
The $\kbol-\Lbol$ relations derived in the present work are calibrated for the first time against a sizable AGN sample, and rely on observed redshifts, X--ray luminosities and column density distributions.
We find that $\kbol$ is significantly lower at high $\Lbol$ with respect to previous estimates by Marconi et al. (2004) and Hopkins et al. (2007).
Black hole masses and Eddington ratios are available for 170 Type-1 AGN, while black hole masses for Type-2 AGN are computed for 481 objects using the black hole mass-stellar mass relation and the morphological information. We confirm a trend between $\kbol$ and $\lambdaEdd$, with lower hard X--ray bolometric corrections at lower Eddington ratios for both Type-1 and Type-2 AGN. We find that, on average, Eddington ratio increases with redshift for all Types of AGN at any given $\Mbh$, while no clear evolution with redshift is seen at any given $\Lbol$.
\end{abstract}

\section{Introduction}
It is widely accepted that the central engine of AGN are accreting supermassive black holes (SMBHs) at the center of galaxies with masses of the order of $10^{6-9}M_\odot$ \citep{1964ApJ...140..796S,1969Natur.223..690L}.
Locally, the SMBH mass correlates with the mass of the bulge of the host-galaxy (\citealt{1998AJ....115.2285M,2003ApJ...589L..21M}), with the velocity dispersion of the bulge (\citealt{2000ApJ...539L...9F,2002ApJ...574..740T}), and with the luminosity of the bulge (\citealt{1995ARA&A..33..581K}).
\rev{The existence of these correlations implies that the growth of the SMBH is tightly linked with the galaxy evolution, playing a crucial role in the star-formation history of the galaxy itself. 
The feedback between SMBH and host-galaxy is therefore a pivotal ingredient that has to be taken into account in SMBH/galaxy formation and co-evolution studies (see \citealt{1998A&A...331L...1S,1999MNRAS.308L..39F,2001ApJ...552L..13C,2005Natur.433..604D}).}
A fundamental question is how the AGN energy detected as radiation is produced. 
AGN are a broad-band phenomenon, hence a multi-wavelength approach is mandatory in order to understand the physics that underlie the AGN emission.
Several studies have been performed on the shape of the SED, as parameterized by the correlation between the $\alphaox$ index, \rev{defined as $\alphaox=-\Log[\lx/\lo]/2.605$},  and the optical luminosity (\citealt{avnitananbaum79,zamorani81,vignali03,steffen06,just07,2009ApJS..183...17Y,2010A&A...512A..34L,2012A&A...539A..48M}), or between $\alphaox$ and the Eddington ratio \citep{vasudevanfabian07,kelly08,2009MNRAS.399.1553V,2009MNRAS.392.1124V,2011ApJ...733...60T}. 
How $\alphaox$ evolves with luminosity may provide a first hint about the nature of the dominant energy generation mechanism in AGN. It is also a first step towards an estimate of the AGN bolometric luminosity function (\citealt{hopkins07}, hereafter H07; \citealt{2009ApJ...690...20S}) and the mass function of SMBHs (\citealt{2004MNRAS.354.1020S,marconi04}, hereafter M04).
All these works consider a model intrinsic quasar SED, which is described with a series of broken power-laws, similar to those objects in bright optically selected samples \citep{1994ApJS...95....1E,2006ApJS..166..470R}. M04 and H07 have also studied the relationship between the bolometric correction, $\kbol$, as a function of the bolometric luminosity, $\Lbol$, in different bands (e.g., the B-band at 0.44$\mu$m, the soft and the hard X--ray bands at [0.5-2] keV and [2-10] keV, respectively). 
Therefore, it is of fundamental importance to verify these correlations considering statistically relevant samples of both broad-line (Type-1) and narrow-line (Type-2\footnote{In the following we will adopt the nomenclature ``Type-2'' AGN, ``not Type-1'', or ``not broad-line'' AGN for the same population of sources.}) AGN over a wide range of redshift and luminosities. 
\par
We analyze the dependence of $\kbol$ on $\Lbol$ in the B-band, and in the soft and hard X--ray bands using a large X-ray selected sample of both Type-1 and Type-2 AGN in the Cosmic Evolution Survey (COSMOS) field (\citealt{scoville07}) from the XMM-COSMOS sample (\citealt{hasinger07}).
The COSMOS field is a unique area for its deep and wide multi-wavelength coverage: radio with the VLA, infrared with {\it Spitzer} and \textit{Herschel}, optical bands with {\it Hubble}, {\it Subaru}, SDSS and other ground-based telescopes, near- and far-ultraviolet bands with the \textit{Galaxy Evolution Explorer} (GALEX) and X-rays with XMM--{\it Newton} and {\it Chandra}. The spectroscopic coverage with VIMOS/VLT and IMACS/Magellan, coupled with the reliable photometric redshifts derived from multiband fitting (\citealt[herefter S11]{2011arXiv1108.6061S}), allows us to build a large and homogeneous sample of AGN with a well sampled spectral coverage and to keep selection effects under control.
In \citet[hereafter L10]{2010A&A...512A..34L} bolometric corrections and Eddington ratios for hard X--ray selected Type-1 AGN in the COSMOS field are presented. That work showed that the bolometric parameters are useful to give an indication of the accretion rate onto SMBH (i.e., a lower bolometric correction corresponds to a lower Eddington ratio, and vice-versa; see also \citealt{2009MNRAS.392.1124V,2009MNRAS.399.1553V}, hereafter V09a and V09b respectively). However, in L10, V09a and V09b the study was mainly focused on the bolometric output of Type-1 AGN, while an alternative approach is needed in order to study the bolometric parameters and Eddington ratios of Type-2 AGN. \citet{2010MNRAS.402.1081V} (hereafter V10) explore a method for characterizing the bolometric output of both obscured and unobscured AGN by adding nuclear IR and hard X-ray luminosities (as originally proposed by \citealt{2007A&A...468..603P}). They also estimate Eddington ratios using black hole mass estimates from the black hole mass-host galaxy bulge luminosity relation for obscured and unobscured AGN finding a significant minority of higher accretion rate objects amongst high-absorption AGN.
\citet{2011A&A...534A.110L} (hereafter L11) present an SED-fitting method for characterizing the bolometric output of obscured AGN, and giving also an estimate of stellar masses and star formation rates for Type-2 AGN.
In this analysis the mid-infrared emission is utilized as a proxy to constrain the thermal emission, and it is used, in conjunction with the X--ray emission, to give an estimate of the bolometric luminosity. 
Taking advantage of these results, we are able to build up a homogeneous analysis to further study the bolometric output of X--ray selected AGN at different absorption levels. 
Moreover, several improvements are included in this study with respect to L10 and L11. Firstly, the inclusion of \textit{Herschel} data at 100~$\mu$m and 160~$\mu$m (\citealt{2011A&A...532A..90L}) in the SED-fitting code is of fundamental importance to better constrain the far-infrared emission and, therefore, the AGN emission in the mid-infrared for Type-2 AGN. Photometric redshifts are taken into account in order to extend our sample also to fainter magnitudes using the updated release of photometric redshifts provided by S11.
Finally, we have included the H band photometry now available in the COSMOS field; this allows us to further increase the coverage in the near-infrared\footnote{These data were reduced using similar methods as described in \citet{2010ApJ...708..202M}.}.
Black hole masses and Eddington ratios for Type-2 AGN are estimated through scaling relations (\citealt{2004ApJ...604L..89H}) using a Monte Carlo method in order to account for uncertainties in stellar masses, bolometric luminosities, as well as the intrinsic scatter in the $\Mbh-M_\ast$ relation. The morphological information is also taken into account in order to estimate the bulge-to-total luminosity/flux ratio (see \citealt{2011ApJ...734..121S} and Sect.~\ref{Eddington ratios} of the present paper for further details) which is used to further constrain the black hole mass estimate.
\par
This paper is organized as follows. In Sect.~\ref{The Data Set} we report the selection criteria for the hard X--ray selected samples of Type-1 and Type-2 AGN used in this work. Section~\ref{Rest-frame monochromatic fluxes and Spectral Energy Distributions} presents the multi-wavelength data-set, while Section~\ref{Bolometric luminosities and bolometric corrections} describes the methods used to compute intrinsic bolometric luminosites and bolometric corrections for Type-1 and Type-2 AGN. Section~\ref{Eddington ratios} describes black hole mass estimates and Eddington ratios, while in Sect.~\ref{Results and discussion} we discuss our findings. In Sect.~\ref{Summary and Conclusions} we summarize the most important results.
\par
We adopted a flat model of the universe with a Hubble constant $H_{0}=70\, \rm{km \,s^{-1}\, Mpc^{-1}}$, $\Omega_{M}=0.27$, $\Omega_{\Lambda}=1-\Omega_{M}$ (\citealt{komatsu09}).

\begin{table}
\caption{Summary of the photometric classification criteria.\label{tbl-selection}} 
\centering
\begin{tabular}{@{}l c|l l}
Sample & Total & SED-Type$^{\mathrm{a}}$ & N$^{\mathrm{b}}$ \\ [0.5ex]
       &   &  &  \\ [0.5ex]
\hline\hline\noalign{\smallskip}
                &     & Galaxy-dominated SED & \textbf{263}  \\[0.3ex]
Photo-$z$       & 330 &  &  \\[0.3ex]
                &     & AGN-dominated SED & \textbf{67}  \\[0.3ex]
\hline\noalign{\smallskip}
                   &     & Galaxy-dominated SED & 39 (12\%) \\[0.3ex]
Type-1 Spectro-$z$ & \textbf{315} & AGN-dominated SED & 271\\[0.3ex]
                   &     & no photo-$z$ Class & 5\\[0.3ex]
\hline\noalign{\smallskip}
                   &     & Galaxy-dominated SED & 257\\[0.3ex]
Type-2 Spectro-$z$ & \textbf{284} & AGN-dominated SED & 25 (9\%) \\[0.3ex]
                   &     & no photo-$z$ Class & 2\\[0.3ex]
\hline\hline
\end{tabular}
\flushleft\begin{list}{}
 \item[${\mathrm{a}}$] Photometric classification according to S11. Templates 19-30 have AGN-dominated SEDs, while templates 1-18 and 101-130 have galaxy-dominated SEDs.
 \item[][${\mathrm{b}}$] Number of sources in each sub-sample: galaxy-dominated SEDs, AGN-dominated SEDs, and sources without photometric classification.
\end{list}
\end{table}

\section{The Data Set}
\label{The Data Set}
The Type-1 and Type-2 AGN samples discussed in this paper are extracted from the XMM-COSMOS catalog which comprises $1822$ point--like X--ray sources detected by XMM-\textit{Newton} over an area of $\sim 2~\rm deg^2$ \citep{hasinger07,cappelluti09}.
All the details about the catalog are reported in \cite{2010ApJ...716..348B}.
We consider in this analysis 1577 X--ray selected sources for which a reliable optical counterpart can be associated (see discussion in \citealt{2010ApJ...716..348B}, Table 1)\footnote{The multi-wavelength XMM-COSMOS catalog can be retrieved from: http://www.mpe.mpg.de/XMMCosmos/xmm53\_release/, version $1^{\rm st}$ November 2011.}.
A not negligible fraction of AGN may be present among optically faint X--ray sources without optical spectroscopy; therefore, in order to extend both Type-1 and Type-2 AGN samples to fainter luminosities, we have considered the updated release of the photometric catalog provided by S11. Photometric classification from 1 to 30 include galaxy-dominated SEDs (early type, late type and ULIRG galaxies), low-and-high luminosity AGN SEDs and hybrids created assuming a varying ratio between AGN and galaxy templates (see Table~2 in \citealt{salvato09} for details). Sources coded from 100 to 130 are reproduced by host-galaxy dominated SED (elliptical, spiral and star-forming galaxies, see Fig.~1 in \citealt{ilbert09}).
We differentiate photometric classification using the best-fit template, separating those dominated by AGN emission and galaxy emission.
Selection criteria for the Type-1 and the Type-2 AGN samples are described in the following Section.

\begin{figure*}
  \centering
  {\includegraphics[width=0.3\textwidth]{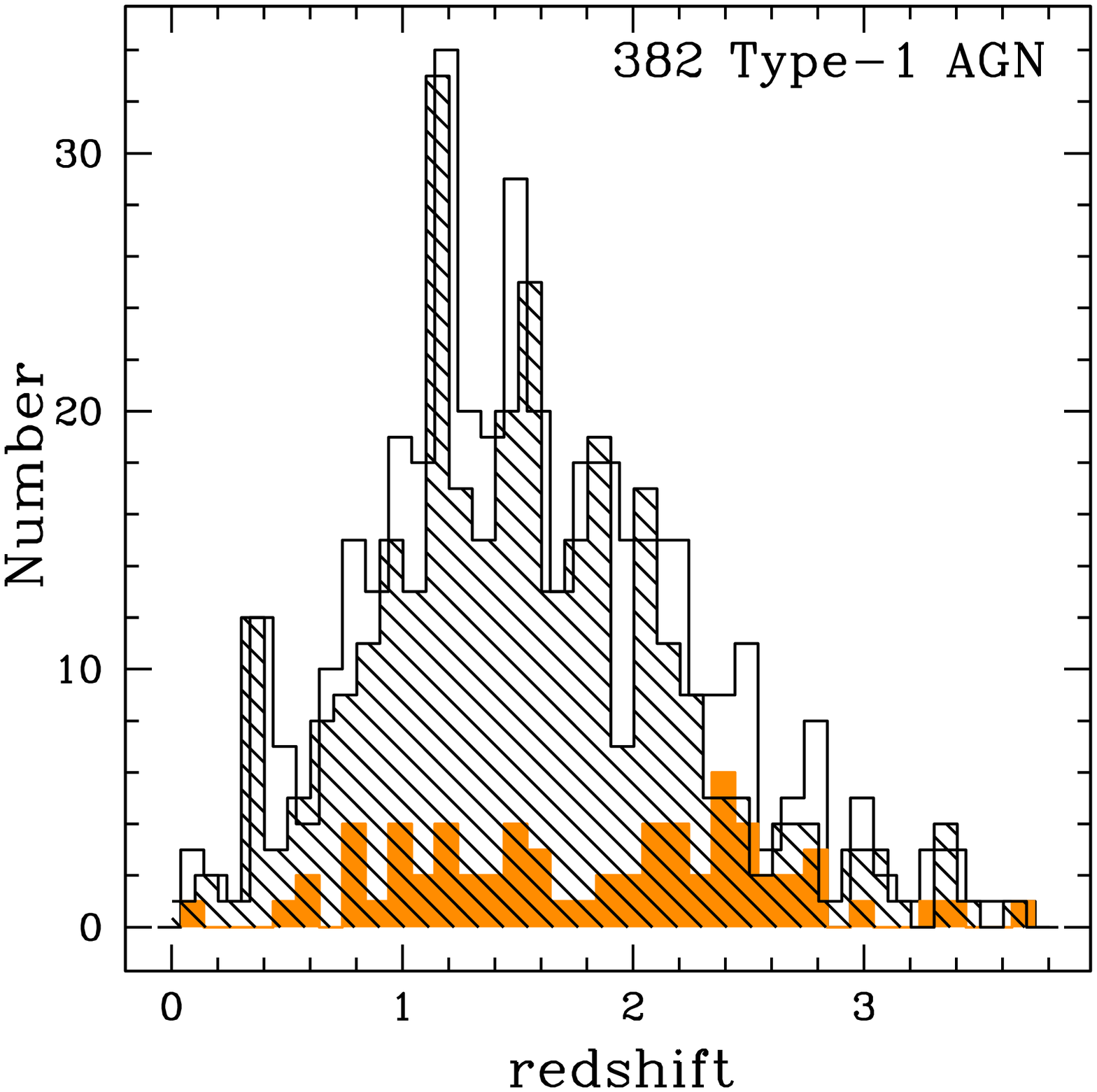}}
  {\includegraphics[width=0.3\textwidth]{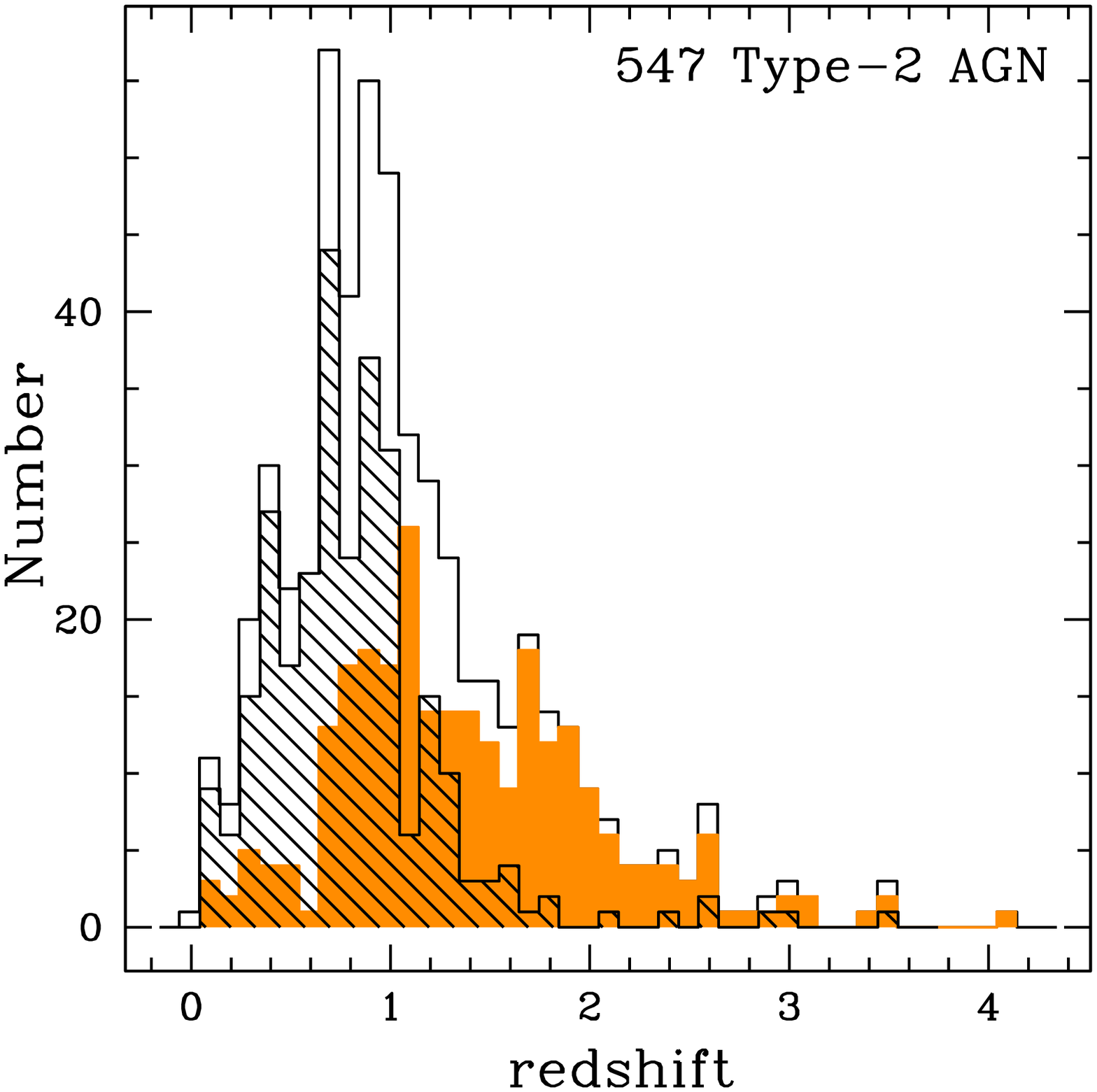}}                        
  \caption{Redshift distribution of the hard X--ray selected Type-1 (\textit{left panel}) and Type-2 (\textit{right panel}) AGN samples considered in this work. The hatched histogram shows the redshift distribution for the sample of spectroscopically identified sources, while the filled histogram is the redshift distribution for the sources without spectroscopic redshift. The total sample is reported with the open histogram.}
  \label{fig:histz}
\end{figure*}

\subsection{Type-1 and Type-2 AGN samples}
\label{Type-12 sample}
We have selected 971 X--ray sources detected in the [2-10]~keV band at a flux larger than $3\times10^{-15}\fluxunits$ (see \citealt{2010ApJ...716..348B}). From this sample, 315 objects are spectroscopically classified broad-line AGN on the basis of broad emission lines ($FWHM > 2000 \,{\rm km \; sec^{-1}}$) in their optical spectra (see \citealt{lilly07,trump09}). 
We will refer to this sample as the ``spectro-z" Type-1 AGN sample. We consider Type-2 AGN all the rest, including Seyfert 2 AGN, emission-line galaxies, and absorption-line galaxies. 
The remainder 284 AGN with optical spectroscopy do not show broad emission lines (FWHM$<2000$ km s$^{-1}$) in their optical spectra: 254 are objects with either unresolved, high-ionization emission lines, exhibiting line ratios indicating AGN activity,  or not detected high-ionization lines, where the observed spectral range does not allow to construct line diagnostics. Thirty are classified absorption-line galaxies, i.e., sources consistent with a typical galaxy spectrum showing only absorption lines. 
We will refer to this sample as the ``spectro-z" Type-2 AGN sample.
\par
In order to extend our Type-1 and Type-2 AGN samples to fainter magnitudes, we proceed as follow.
We have selected all sources with a best-fit photometric classification consistent with an AGN-dominated SED (i.e., $19\leq{\rm SED-Type}\leq30$; see S11, for details). In the following, we assume that 67 X-ray sources, classified by the SED fitting with an AGN-dominated SED are Type-1 AGN.
We will refer to this sample as the ``photo-z" Type-1 AGN sample.
From the total photo-z sample we additionally selected all AGN with a best-fit photometric classification inconsistent with a broad-line AGN SED.
We will refer to this sample (i.e., 263 X-ray sources with ${\rm SED-Type}>100 ~\&~ {\rm SED-Type}<19$, see Table~\ref{tbl-selection}) as the ``photo-z" Type-2 AGN sample. 
\par
The final Type-1 AGN sample used in our analysis comprises 382 X-ray selected AGN (315 from the spectro-z sample and 67 from the photo-z sample), while the final Type-2 AGN sample comprises 547 X-ray selected AGN (284 from the spectro-z sample and 263 from the photo-z sample). These samples span a wide range of redshifts and X-ray luminosities.
Only 12 Type-2 AGN (2\%, 9 and 3 AGN from the spectro-z and photo-z sample, respectively) have $\Log\Lhard<42$ erg s$^{-1}$. Sources with $\Lhard<42$ might be interpreted as star-forming galaxies, however the number of these objects is very small and their inclusion in the Type-2 class does not affect any of the results.
\par
We have estimated the expected contamination and incompleteness in the classification method, on the total Type-1 and Type- 2 AGN samples, by checking the distribution of the photometric classification for the spectroscopically identified Type-1 and Type- 2 AGN samples. The large majority of the broad emission line AGN in the spectro-z sample are classified as Type-1 AGN by the SED fitting as well (271/315; 86\%),  while the number of spectroscopic Type-1 AGN which have SED--Type $>$ 100 or SED--Type $<$ 19 is relatively small (39 sources). Similarly, a good agreement between the two classifications is present also for the Type-2 AGN; for 91\% (257/284)  of the 284 Type-2 sources with spectroscopic redshift, the spectroscopic classification is in agreement with the SED-based classification.  If we make the reasonable assumption that the fraction of agreement between the two classifications is the same also for the sample for which we do not have a spectroscopic classification, we find that our total Type-1 AGN sample is expected to be contaminated (i.e. Type-2 AGN misclassified as Type 1 AGN from the SED analysis) at the level of $\sim$1.6\% (6 objects) and incomplete (i.e. Type-1 AGN misclassified as Type-2 AGN, and therefore not included in our Type-1 sample) at the level of $\sim$9.2\%.  The same fractions for the total Type-2 sample are $\sim$6.4\% (contamination) and $\sim$1.1\% (incompleteness). 
In Table~\ref{tbl-selection} a summary of the selection criteria and the numbers of AGN in the various classes are presented.
\par
The redshift distributions of the total, spectroscopic and photometric Type-1 and Type-2 AGN samples are presented in Figure~\ref{fig:histz} (left panel). The median redshift of the total Type-1 AGN sample is 1.51 (the mean redshift is 1.57, with a dispersion of 0.70). The median redshift of the spectro-z sample is 1.45, while the median redshift of the photo-z sample is 1.92.
The median redshift of the total Type-2 AGN sample is 0.96 (the mean redshift is 1.10, with a dispersion of 0.64). The median redshift of the spectro-z sample is 0.81, while the median redshift of the photo-z sample is 1.41.

\begin{figure*}
  \centering
  {\includegraphics[width=0.3\textwidth]{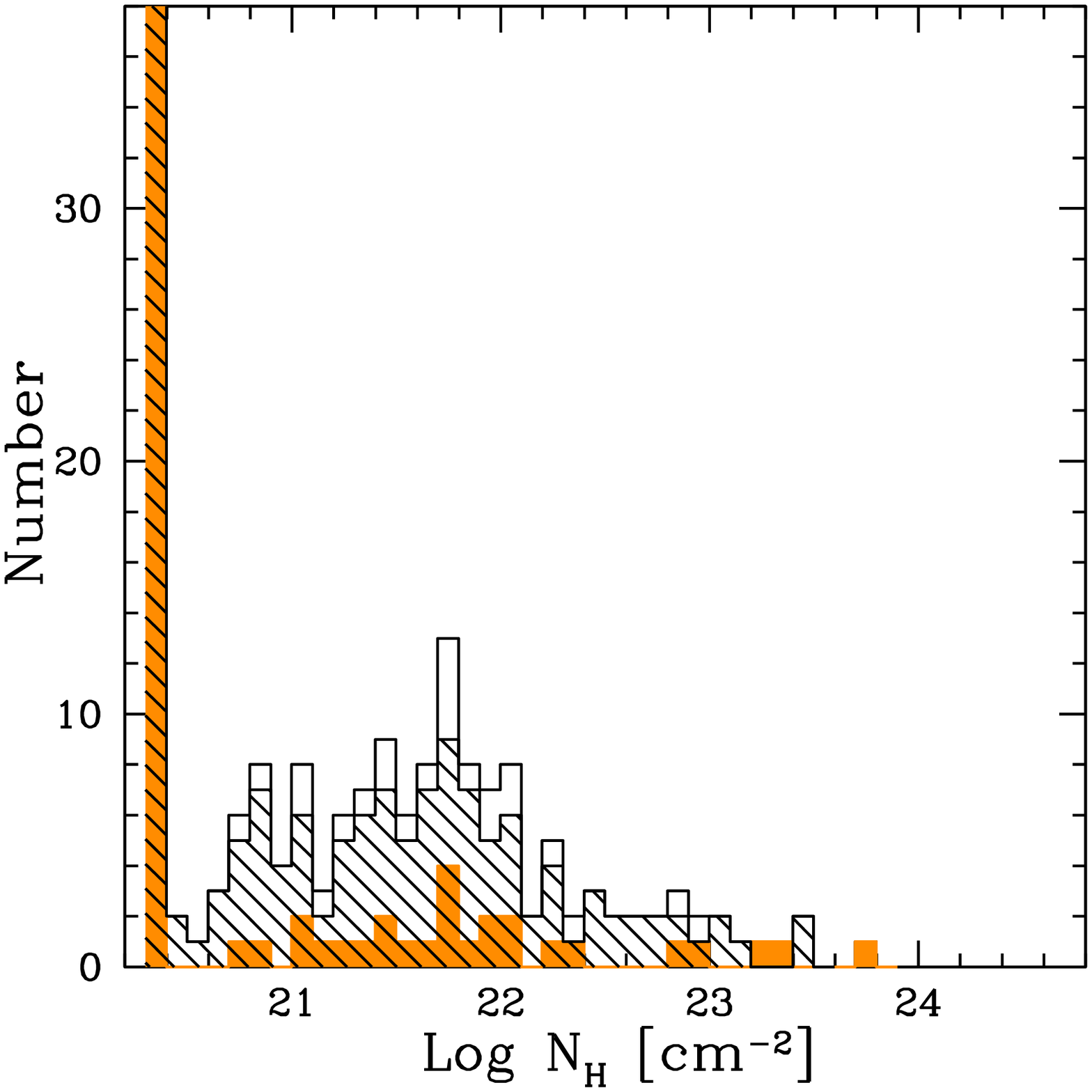}}
  {\includegraphics[width=0.3\textwidth]{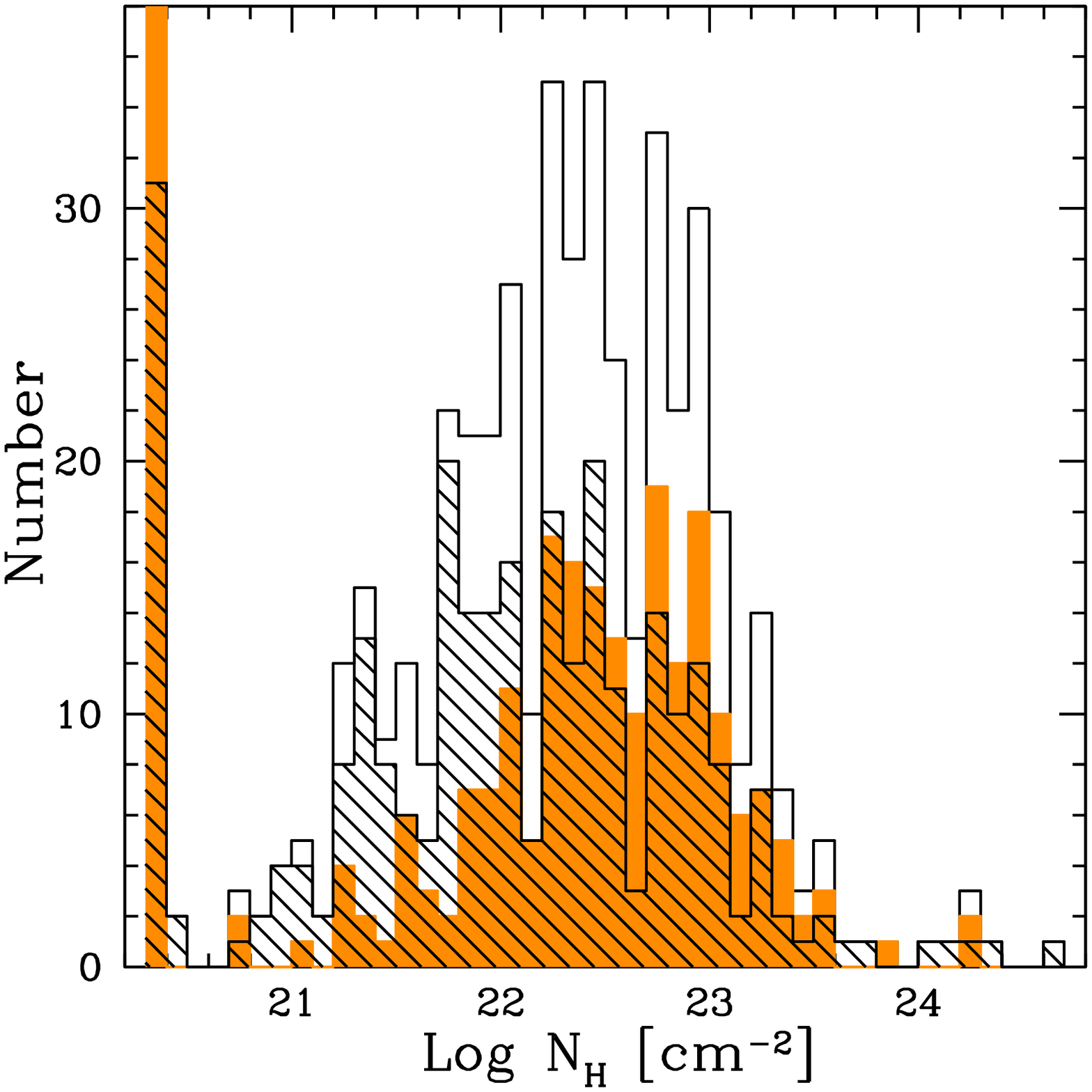}}                        
  \caption{Column density distribution of the Type-1 (\textit{left panel}) and Type-2 (\textit{right panel}) AGN samples (\textit{black open histogram}). About 67\% of the Type-1 AGN sample and 16\% of the Type-2 AGN sample have $\NH$ values consistent with the Galactic one. Note that the first bin of in the left panel is much higher than what is plotted in the histogram (206 spectro-z and 40 photo-z have $\NH=20.5$  cm$^{-2}$), while for the right panel 56 spectro-z and 33 photo-z have $\NH$ consistent with the Galactic value. The hatched histogram shows the $\NH$ distribution for the sample of spectroscopically identified sources, while the filled histogram is the $\NH$ distribution for the sources without spectroscopic redshift. Sixty percent (322/547) of the Type-2 AGN sample and 10\% (39/382) of the Type-1 AGN sample have $\Log \NH\geq22$ cm$^{-2}$.}
  \label{fig:histnh}
\end{figure*}

\section{Spectral Energy Distributions}
\label{Rest-frame monochromatic fluxes and Spectral Energy Distributions}
We have collected the multi-wavelength information from mid-infrared to hard X--rays as in L10 and L11.
The observations in the various bands are not simultaneous, as they span a time interval of about 5 years: 2001 (SDSS), 2004 (Subaru and CFHT) and 2006 (IRAC). 
Variability for absorbed sources is likely to be a negligible effect, which is probably not the case for Type-1 AGN. Therefore, in order to reduce variability effects, we have selected the bands closest in time to the IRAC observations (i.e., we excluded SDSS data, that in any case are less deep than other data available in similar bands).
All the data for the SED computation were shifted to the rest frame, so that no K-corrections were needed.
Galactic reddening has been taken into account: we used the selective attenuation of the stellar continuum $k(\lambda)$ taken from Table 11 of \cite{capak07}. Galactic extinction is estimated from \cite{schlegel98} for each object. 
We decided to consider the near-UV GALEX band for Type-1 and Type-2 AGN with redshift lower than 1, and far-UV GALEX band for sources with redshift lower than 0.3 in order to avoid Ly$\alpha$ absorption from foreground structures. 
In the far-infrared the inclusion of \textit{Herschel} data at 100~$\mu$m and 160~$\mu$m (\citealt{2011A&A...532A..90L}) better constrain the AGN emission in the mid-infrared. The number of detections at 100~$\mu$m is 63 (16\%, 59 spectro-z and 4 photo-z) for the Type-1 AGN sample, while is 98 (18\%, 73 spectra-z and 25 photo-z) for the Type-2 AGN sample.
At 160~$\mu$m the number of detections for the Type-1 AGN sample is 56 (15\%, 52 spectro-z and 4 photo-z), while is 87 (16\%, 63 spectra-z and 24 photo-z) for the Type-2 AGN sample. 
Count rates in the 0.5-2 keV and 2-10 keV are converted into monochromatic X--ray fluxes in the observed frame at 1 and 4 keV, respectively, using a Galactic column density $\NH = 2.5 \times 10^{20}\,cm^{-2}$ (see \citealt{1990ARA&A..28..215D,2005A&A...440..775K}).
We have computed the integrated unabsorbed luminosity in the [0.5-2]keV and [2-10]keV bands for both Type-1 and Type-2 AGN samples.
For a sub-sample of 100 Type-1 AGN (26\%) and 240 Type-2 AGN (44\%) we have an estimate of the column density $N_{\rm H}$ from spectral analysis (see \citealt{2007ApJS..172..368M,2010A&A...514A..85M}), while for 282 Type-1 AGN and 307 Type-2 AGN absorption is estimated from hardness ratios (HR; see \citealt{2010ApJ...716..348B}). 
The integrated intrinsic unabsorbed luminosity is computed assuming a power-law spectrum with slope $\Gamma=2$ and $\Gamma=1.7$ for the [0.5-2]keV and [2-10]keV bands, respectively \citep{cappelluti09}.
\par
\begin{figure*}
 \includegraphics[width=12cm,clip]{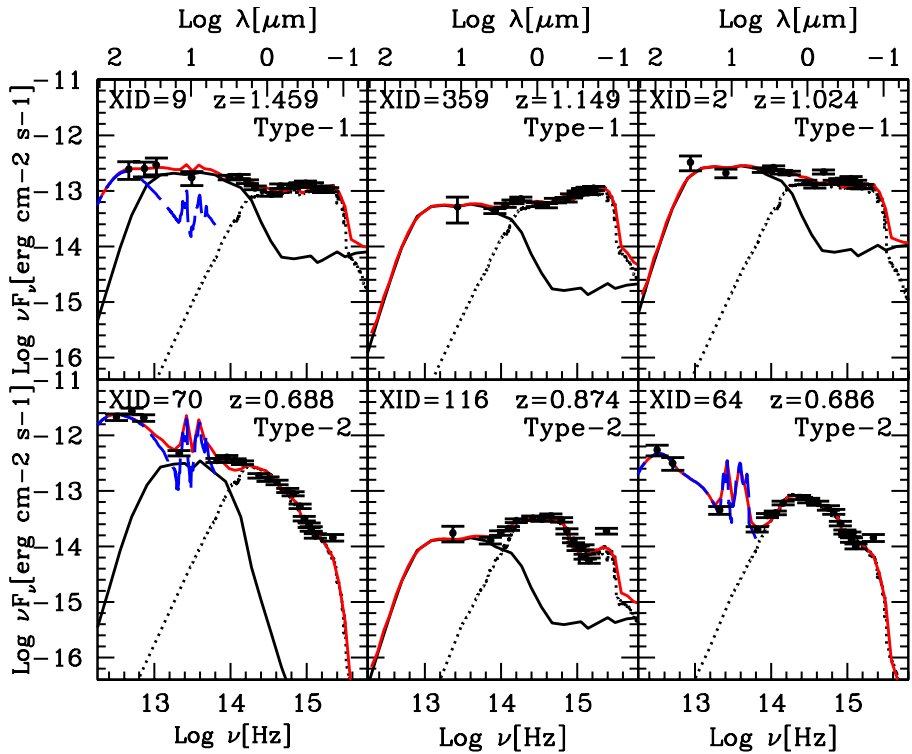}
 \caption{Examples of SED decompositions. Black circles are the observed photometry in the rest-frame (from the far-infrared to the optical-UV). The blue long-dashed, black solid and dotted lines correspond respectively to the starburst, AGN and host-galaxy templates found as the best fit solution. The red line represents the best-fit SED. The top three panels are Type-1 AGN, while the bottom three are Type-2 AGN.}
 \label{panel}
\end{figure*}
In Figure \ref{fig:histnh} we show the distribution of column densities for the Type-1 and Type-2 AGN samples. Two-hundred and twenty-four Type-1 AGN ($\sim64\%$) and 87 Type-2 AGN ($\sim16\%$) do not require absorption in addition to the Galactic one.
The mean $N_{\rm H}$ value is $7.4\times10^{20}$ cm$^{-2}$ for the Type-1 AGN and $\sim10^{22}$ cm$^{-2}$ for the Type-2 AGN. Sixty percent (322/547) of the Type-2 AGN sample and 10\% (39/382) of the Type-1 AGN sample have $\Log \NH\geq22$ cm$^{-2}$.
The average shift induced by the correction for absorption in the Type-1 sample is, as expected, small in the soft band $\langle\Delta \Log\Lsoft\rangle=0.10\pm0.01$, negligible in the hard band.
The same shift in the Type-2 sample is $\langle\Delta \Log\Lsoft\rangle=0.30\pm0.04$ in the soft band, while it is $\langle\Delta \Log\Lhard\rangle=0.033\pm0.004$ for the hard band.
\par
We have computed the individual rest-frame SEDs for all sources in the sample, following the same approach as in L10.
For the computation of the bolometric luminosity for Type-1 AGN we need to extrapolate the UV data to X-ray gap and at high X-ray energies. We have extrapolated the SED up to 1200$\text{\AA}$ with the slope computed using the last two rest-frame optical luminosity data points at the highest frequency in each SED (only when the last optical-UV rest frame data point is at $\lambda>1200\AA$). Then, we assume a power law spectrum to 500$\text{\AA}$, as measured by HST observations for radio-quiet AGN ($f_\nu\propto \nu^{-1.8}$, see \citealt{zheng97}). We then linearly connect (in the log space) the UV luminosity at 500 $\text{\AA}$ to the luminosity corresponding to the frequency of 1 keV.

\section{Bolometric luminosities}
\label{Bolometric luminosities and bolometric corrections}
\rev{The mid-infrared luminosity is considered an indirect probe of the accretion disk optical/UV luminosity (see \citealt{2007A&A...468..603P,2010A&A...517A..11P,2010MNRAS.402.1081V})}.
The nuclear bolometric luminosity for Type-2 AGN is then estimated by using the same approach as in L11, whereas the sum of the infrared and X--ray luminosities are used as a proxy for the intrinsic nuclear luminosity ($\Lbol=\Lir+\Lx$). 
The main purpose of the SED-fitting code is to disentangle the various contributions (star-burst, AGN,
host-galaxy emission) in the observed SEDs by using a standard $\chi^2$ minimization procedure.
The code is based on a large set of star-burst templates from \citet{2001ApJ...556..562C} and \citet{2002ApJ...576..159D}, and galaxy templates from the \citet{2003MNRAS.344.1000B} code for spectral synthesis models, while AGN templates are taken from \citet{2004MNRAS.355..973S}. These templates represent a wide range of SED shapes and luminosities and are widely used in the literature. 
After performing the SED-fitting, only the nuclear component of the best-fit is integrated. Hence, the total IR luminosity $\Lir$ is obtained integrating the nuclear template between 1 and 1000~$\mu$m. To convert this IR luminosity into the nuclear accretion disk luminosity, we applied the correction factors to account for the torus geometry and the anisotropy ($\sim1.7$, see \citealt{2007A&A...468..603P,2010A&A...517A..11P}).
The total X--ray luminosity $\Lx$ is estimated by integrating the X--ray SED in the 0.5-100 keV range. 
\par
The photometric data used in the SED-fitting code, from low to high frequency, are: \textit{Herschel}-PACS bands (160~$\mu$m and 100~$\mu$m), MIPS/Spitzer (24~$\mu$m and 70~$\mu$m), 4 IRAC bands (8.0~$\mu$m, 5.8~$\mu$m, 4.5~$\mu$m and 3.6~$\mu$m), $K_S$ CFHT, $H$ CFHT, $J$ UKIRT, optical broad-band Subaru ($B_J$, $V$, $g^+$, $r^+$ and $z^+$) and CFHT ($i^*$, $u^*$) bands.  
The starburst component is used only when the source is detected at wavelengths longer than 24~$\mu$m rest-frame. Otherwise, a two components SED-fit is used. Eighteen is the maximum number of bands adopted in the SED-fitting (only detection are considered). Figure~\ref{panel} shows the multi-wavelength photometry and the model fits for six AGN, three Type-1 AGN in the top panels and three Type-2 AGN in the bottom panels. The three components adopted in the SED-fitting code, starburst, AGN torus and host-galaxy templates, are shown as a blue long-dashed line, black solid line and dotted line, respectively. 
\par
The bolometric luminosity for Type-1 AGN is usually computed by integrating the observed SED, as described in Sect.~\ref{Rest-frame monochromatic fluxes and Spectral Energy Distributions}, in the $\Log\nu-\Log(\nu L_\nu)$ rest-frame plane from $1\mu$m to 200 keV. The choice to neglect the infrared bump is motivated by the fact that nearly all photons emitted at these wavelengths by the AGN are reprocessed optical/UV/soft X-ray photons; in this way we avoid to count twice the emission reprocessed by dust (see M04).
However, given that we want to compare bolometric parameters for Type-1 and Type-2 AGN, we have decided to use the SED-fitting code above described to compute bolometric luminosities for both samples. 
\par
\begin{figure}
 \includegraphics[width=8cm]{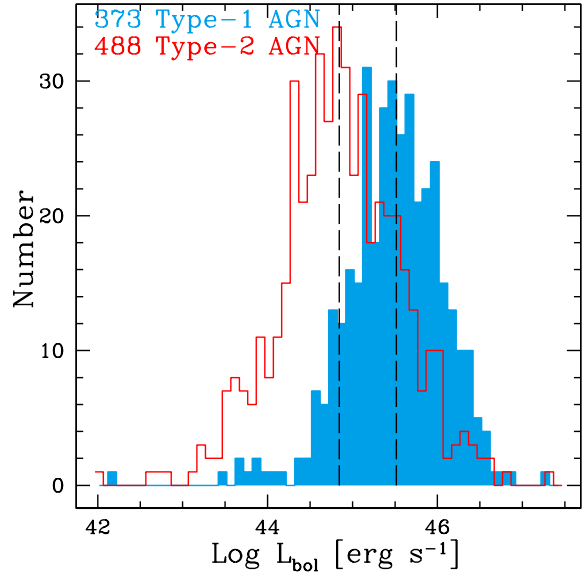}
 \caption{Histogram of bolometric luminosities for Type-1 AGN (\textit{blue filled histogram}), and for Type-2 AGN (\textit{red open histogram}) with AGN best-fit available from the SED-fitting code. The two dashed lines show the median $\Lbol$ values for Type-1 AGN $\Lbol=3.2\times10^{45}$ erg s$^{{-1}}$ (right), and for Type-2 AGN $\Lbol=7.1\times10^{44}$ erg s$^{{-1}}$ (left). }
 \label{histlbol12}
\end{figure}

In order to compute the bolometric correction we used the standard definition
\begin{equation}
 k_{\rm bol,band}=\frac{\Lbol}{L_{\rm band}}
\end{equation}
where $L_{\rm band}$ is the luminosity in soft and hard bands and in the B-band at 0.44$\mu$m. The luminosity in the B-band is computed only for Type-1 AGN, since for Type-2 AGN the emission in the optical is mainly from the host-galaxy.
Bolometric luminosities for Type-1 and Type-2 AGN samples are reported in Figure~\ref{histlbol12}. 
The mean bolometric luminosity value for Type-1 AGN is $\langle\Log \Lbol\rangle=45.50$ with a dispersion of 0.58, while for Type-2 AGN $\langle\Log \Lbol\rangle=44.85$ with a dispersion of 0.70.
Using the SED-fitting approach we found that 9 Type-1 AGN are best-fitted either with only galaxy template (7 objects mostly at redshift higher than 2), or with galaxy plus \rev{star-burst template} (2 objects, all of them have Herschel data at 100/160~$\mu$m). Therefore, bolometric AGN luminosities are not available for these sources.  
For 59 Type-2 AGN (10\% of the main sample) the SED-fitting code is not able to fit the mid-IR part of the SED with the AGN component. 
Twenty-two of these 59 Type-2 AGN are well fitted only with a galaxy component. This sample is predominantly at high redshift,  highly obscured with $\NH>10^{23}$ cm$^{-2}$ and none of these sources have a detection in the far-IR, most of them neither at 24~$\mu$m band. Given that they are mainly at high redshift, all bands are shifted towards high frequencies, therefore the torus component is not needed.
For the remaining 37 Type-2 AGN the best-fit is composed by the galaxy component in the optical and the star-burst component in the far-IR. All these sources present MIPS detection at 24/70~$\mu$m and/or Herschel data at 100/160~$\mu$m. 
These Type-2 AGN are at relatively moderate redshift ($z<1.5$) and X--ray luminosities in the range $40.5\lesssim\Lhard\lesssim44$.
Summarizing, the SED-fitting for 2\% of the Type-1 AGN sample and for 11\% of the Type-2 AGN sample is not able to recover the AGN component in the mid-infrared. 
\par
\begin{figure}
 \includegraphics[width=8cm]{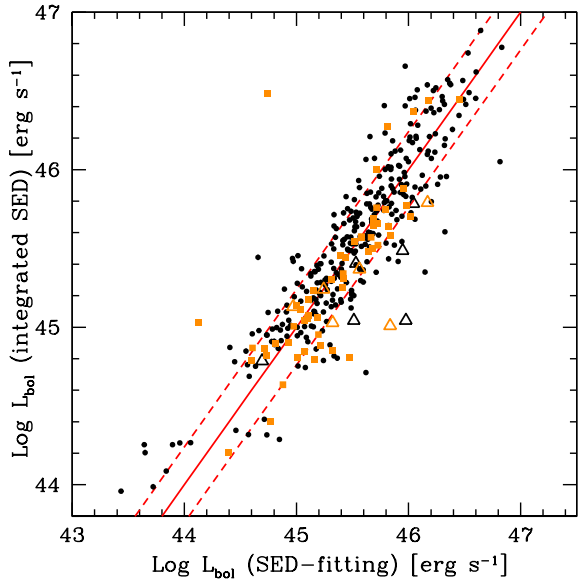}
 \caption{Comparison between the values of bolometric luminosity computed by the SED-fitting code and those obtained integrating the rest-frame SED from 1$\mu$m to the X-ray for the Type-1 AGN sample. Black and orange symbols represent the spectro-z and the photo-z sample respectively. Opens squares represent AGN with an upper limit in the soft X--ray band. The red solid line and the dashed lines represent the one-to-one correlation and 1-$\sigma$ scatter of 0.24 dex, respectively.}
 \label{comparisonlbol}
\end{figure}

We have estimated the uncertainties on bolometric luminosities comparing the bolometric luminosities computed by the SED-fitting code and those obtained integrating the rest-frame SED from 1$\mu$m to the X-ray. Obviously, this test can be only applied to the Type-1 AGN sample. 
In Figure~\ref{comparisonlbol} the comparison between $\Lbol$ from the integrated SED and from the SED-fitting is presented.
We find that the bolometric luminosities are scattered along the one-to-one correlation with a 1-sigma scatter of $\sim0.24$ dex (after performing a 3.5$\sigma$ clipping, 4 objects have been removed). 
Assuming that the errors associated to the two different ways to compute $\Lbol$ are of the same order of magnitude, we can estimate the $1~\sigma$ uncertainty on $\Lbol$ to be $\sim 0.17$ dex.
Since the average uncertainties on the X--ray luminosities are of the order of 10\%, the uncertainty on the bolometric corrections (computed using the soft and hard X--ray luminosities) is dominated by the uncertainty on $\Lbol$ and it is of the order of 0.20 dex.
These values have to be considered as a qualitative indication on the uncertainty on $\Lbol$ and $\kbol$.

\section{Black hole masses and Eddington ratios}
\label{Eddington ratios}

We estimate black hole masses from virial estimators (\citealt{peterson04}) for 170 Type-1 AGN in our sample within the redshift range $0.13 \leq z \leq 3.36$ (mean $\left<z\right> = 1.38$ with a dispersion of 0.49). Of these, 96 use the Mg II line width: 74 sources are from \citet{2010ApJ...708..137M} (with uncertainties of $\sim 0.25$~dex) and 22 from \citet{trump09} (with uncertainties of $\sim 0.4$~dex). The remaining 74 are also from \citet{trump09} and use the H$\beta$ line width, with uncertainties of $\sim 0.4$~dex. We combine these black hole masses with bolometric luminosities to compute the Eddington ratio of each source:
\begin{equation}
\label{eddratiodef} 
 \lambdaEdd=\frac{\Lbol}{\Ledd}\propto\frac{\Lbol}{\Mbh}.
\end{equation}
Virial estimators are unavailable for the Type-2 sample; instead, we exploit the well-studied correlation between black hole and host galaxy bulge mass \citep[in particular that of][]{2004ApJ...604L..89H} to estimate black hole masses. We combine estimates of each host galaxy's total (i.e., bulge+disk) stellar mass from our SED fitting with bulge fractions from morphological assessments in order to determine the stellar mass of the bulge component of each host galaxy. This technique and the uncertainties therein are described below.
\par
We use secure morphological information for 144 Type-2 AGN obtained from the updated Zurich Estimator of Structural Types \citep[ZEST;][]{2007ApJS..172..406S}, known as ZEST+ (Carollo et al. 2012, in preparation). ZEST+ classifies galaxies in five morphological types located in specific regions of the 6-dimensional space. Combining these measured morphologies with the results of extensive AGN host galaxy simulations mapping observed morphology to intrinsic bulge-to-total ratio \citep{2008ApJ...683..644S} yields the following bulge fractions for each ZEST+ morphological type:
\begin{itemize}
\item Elliptical: $\rm{B/Tot} =1$
\item S0: $\rm{B/Tot} =0.5$
\item bulge-dominated disk: $\rm{B/Tot}  =0.75$
\item intermediate-bulge disk: $\rm{B/Tot}=0.5$
\item disk-dominated: $\rm{B/Tot}=0.25$
\end{itemize}
Following \citet{2008ApJ...683..644S} and \citet{2011ApJ...734..121S}, uncertainties in B/Tot are typically $\sim 0.3$, with the limit that $0 \leq \rm{B/Tot} \leq 1$. For 57 spectro-$z$ Type-2 AGN there is significant uncertainty in the morphology (e.g., due to artifacts in the HST-ACS images), while 47 spectro-$z$ objects lie outside the ACS tiles; ZEST+ is therefore unable to determine the morphology of these sources. For these sources and for the photo-$z$ sample, we employed the total stellar mass in the black hole mass estimate (i.e., $\rm{B/Tot} = 1$) and consider the black hole mass as an upper limit.
This does not affect our results, as we will show in Section~\ref{Bolometric correction vs. Eddington ratio}, by treating separately Type-2 AGN with reliable morphological classification.
\par
Uncertainties in the stellar masses are mainly due to two factors. First, each input parameter used to compute $M_*$ has an uncertainty: for example, the metallicity (which we fixed to the solar value), the extinction law (\citealt{2000ApJ...533..682C}), the assumed star formation history, the assumed initial mass function (which is just a scale factor, $\Log M_\ast({\rm Salpeter IMF})\sim \Log M_\ast({\rm Chabrier IMF}) + 0.23$, we have assumed the Chabrier IMF), and different stellar population synthesis models. These inputs carry non-negligible uncertainties and several works in the literature have explored the impact of these uncertainties on the derived physical properties of galaxies \citep[e.g.,][]{2009ApJ...701.1765M, 2009ApJ...699..486C,2010A&A...524A..76B}. The uncertainties on input parameters produce an average scatter of $\sim 0.15$~dex.
\par
A second source of uncertainty is given by the data-set used and it varies for each source. \citet{2010A&A...524A..76B} performed several tests on simulated catalogs and find a scatter of $\sim 0.15$~dex due to this effect. Combining these two sources of uncertainty, we consider a global uncertainty of $\sim 0.2$~dex on the derived values of $M_\ast$.
\par
In order to properly account for these sources of uncertainty in the black hole mass estimates for our Type-2 AGN, we employ a Monte Carlo method, simulating $10^5$ data points for each of our sources within the errors. The method also considers the evolution of the bulge mass-black hole mass relation \citep{2004MNRAS.354L..37M} and the intrinsic scatter in the original relation of \citet{2004ApJ...604L..89H}. We report the median mass of the Monte Carlo distribution for each source and compute the asymmetric uncertainties on the mass based on the distribution.  For the 481 Type-2 AGN with $\Lbol$ available, we further compute the Eddington ratio, additionally accounting for uncertainties in $\Lbol$. Uncertainties in black hole masses and Eddington ratios are slightly asymmetric for those sources with asymmetric uncertainties in B/Tot, but are typically $\sim 0.5$~dex for the black hole masses and $\sim 0.55$~dex for the Eddington ratios. The median black hole mass for the Type-2 AGN sample is $10^8~M_\odot$, while the upper and lower quartiles corresponding to 75\% and 25\% are $2.3\times10^8~M_\odot$ and $4.8\times10^7~M_\odot$, respectively. 
The median black hole mass for the Type-1 AGN sample is $2.7\times10^8~M_\odot$, while the upper and lower quartiles corresponding to 75\% and 25\% are $4.3\times10^8~M_\odot$ and $1.2\times10^8~M_\odot$, respectively. 
\par
\begin{figure*}
  \centering
  {\includegraphics[width=0.3\textwidth]{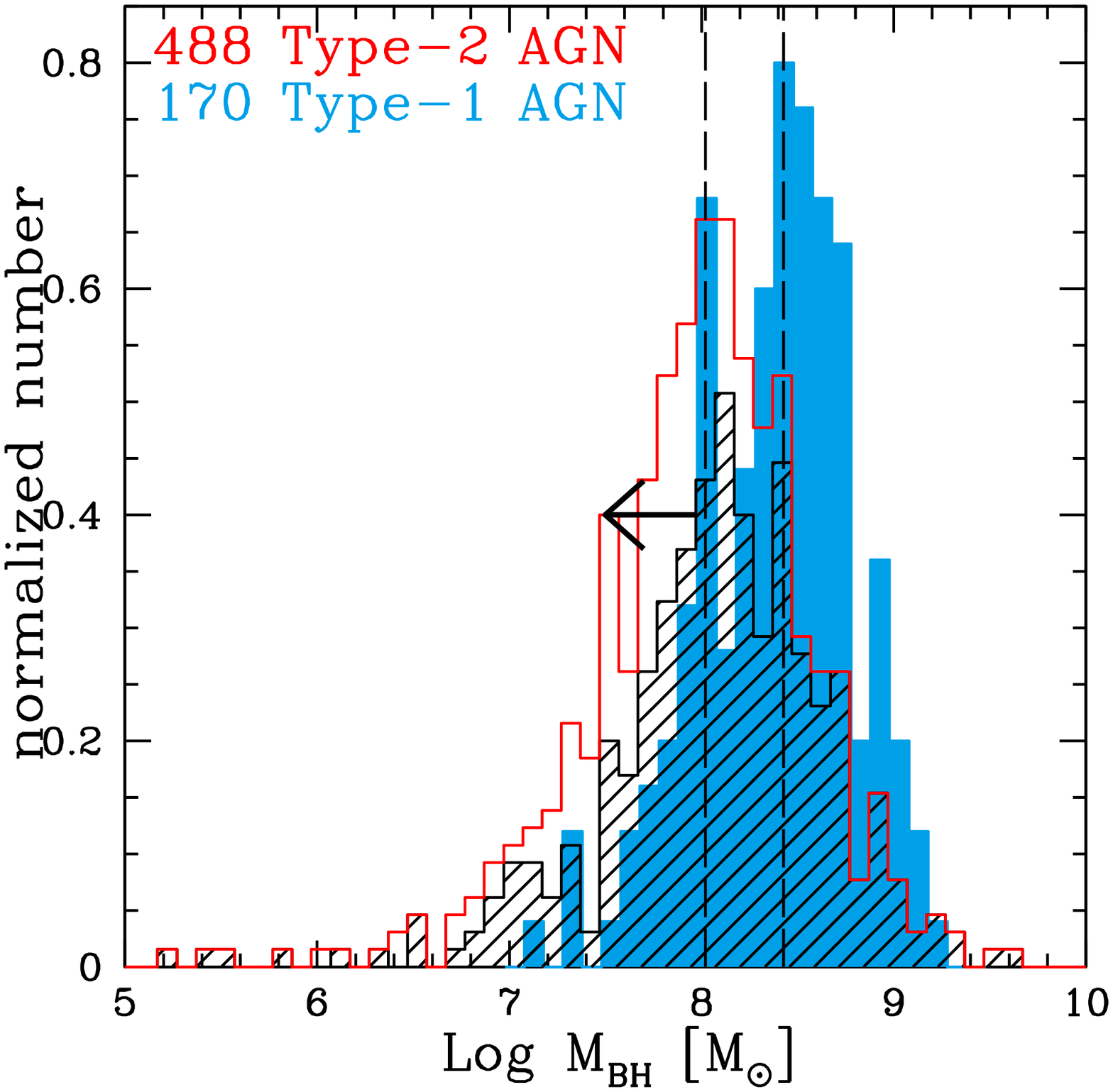}}
  {\includegraphics[width=0.3\textwidth]{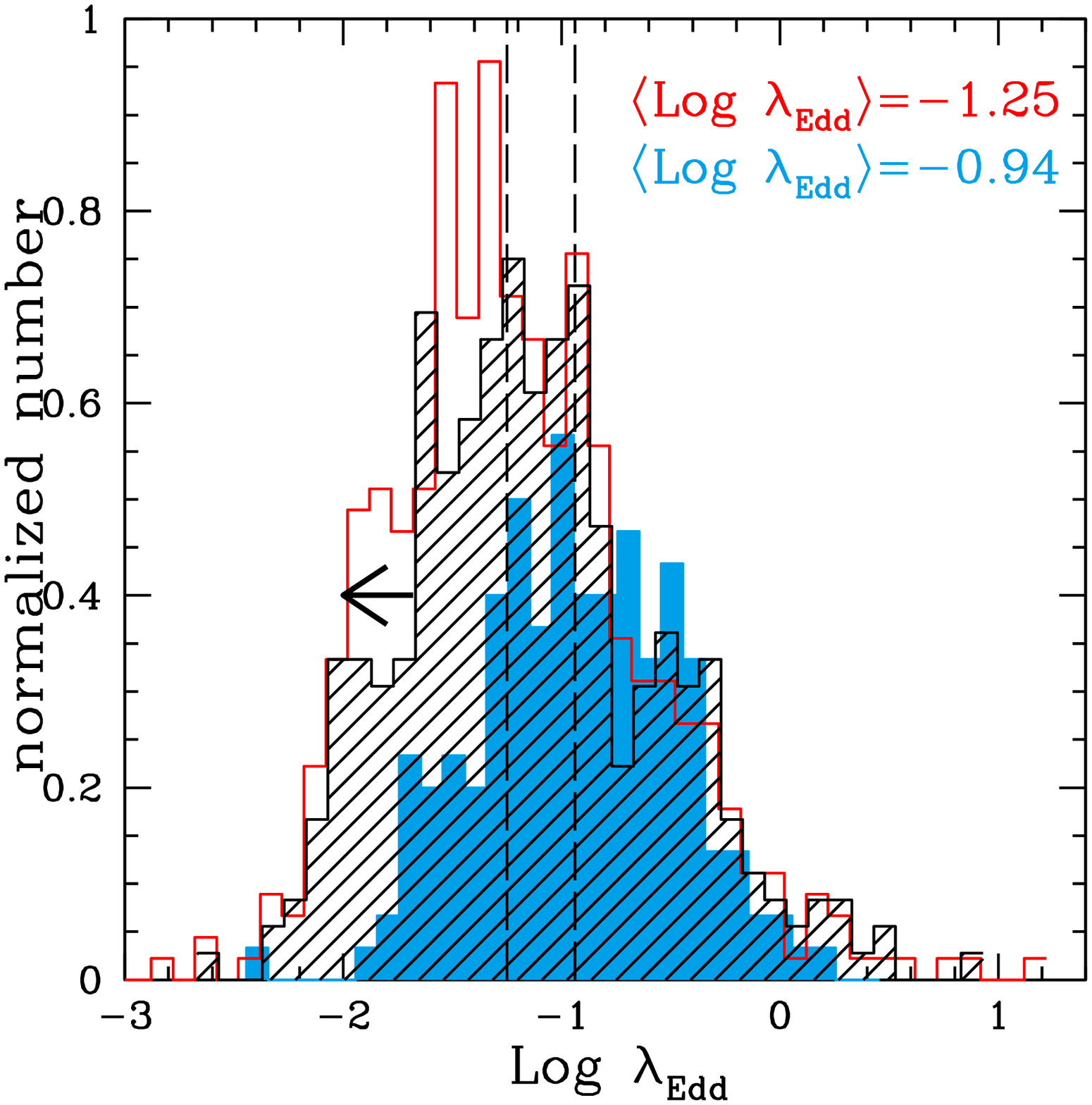}}                        
  \caption{\textit{Left panel}: Histogram of estimated black hole masses from virial estimators for Type-1 AGN (\textit{blue filled histogram}), and from stellar masses for Type-2 AGN (\textit{red open histogram}). The two dashed lines show the median $\Mbh$ values for Type-1 AGN $\Mbh=2.7\times10^8 M_\odot$ (right), and for Type-2 AGN $\Mbh\sim10^8 M_\odot$ (left). \textit{Right panel}: Histogram of estimated Eddington ratios for Type-1 AGN (\textit{blue filled histogram}), and for Type-2 AGN (\textit{red open histogram}). The two dashed lines show the median $\lambdaEdd$ values for Type-1 AGN $\lambdaEdd=0.12$ (right), and for Type-2 AGN $\lambdaEdd=0.06$ (left). Counts are normalized in order to better compare the samples. Black hatched histogram represents the subsample of Type-2 AGN (344 objects) for which B/Tot ratios are fixed to one and $\Mbh$ estimates are considered upper limits.}
  \label{fig:histmbhledd}
\end{figure*}

We show black hole masses and Eddington ratios for the Type-1 and Type-2 AGN samples in Figure~\ref{fig:histmbhledd}. Note that black hole masses for Type 2 AGN would increase by a factor of $\sim 1.7$ using a Salpeter IMF, decreasing the difference between the average Type-1 and Type-2 black hole masses by a similar factor.
For an extensive discussion and comparison between Type-1 and Type-2 AGN see Sects.~\ref{Bolometric correction vs. Eddington ratio}, \ref{Luminosity-redshift dependence} and \ref{Black hole mass-redshift dependence}.

\section{Results}
\label{Results and discussion}
\begin{figure*}
  \centering
  {\label{fig:kbollbolsoft1}\includegraphics[width=0.35\textwidth]{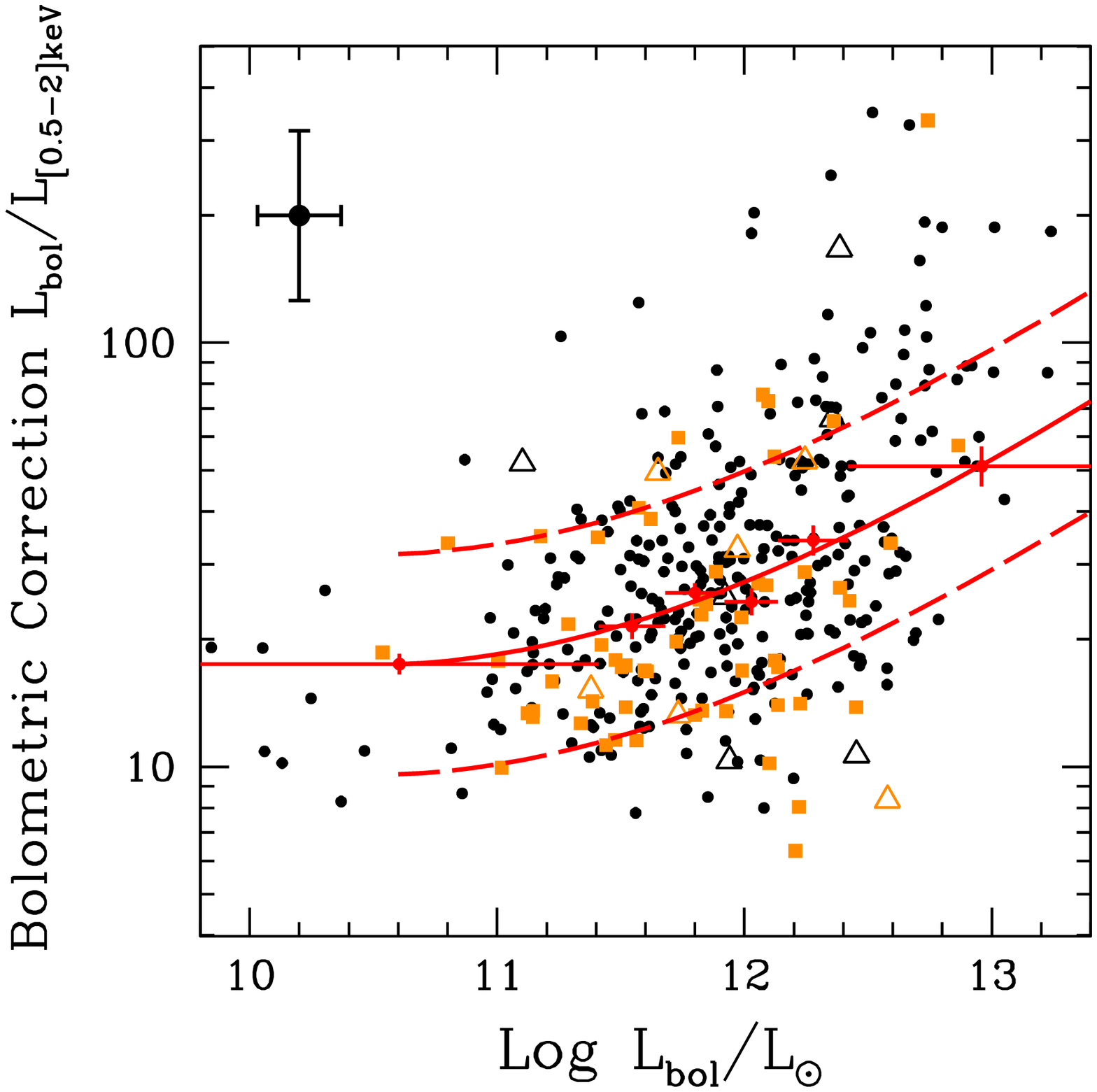}}
  {\label{fig:kbollbolhard1}\includegraphics[width=0.35\textwidth]{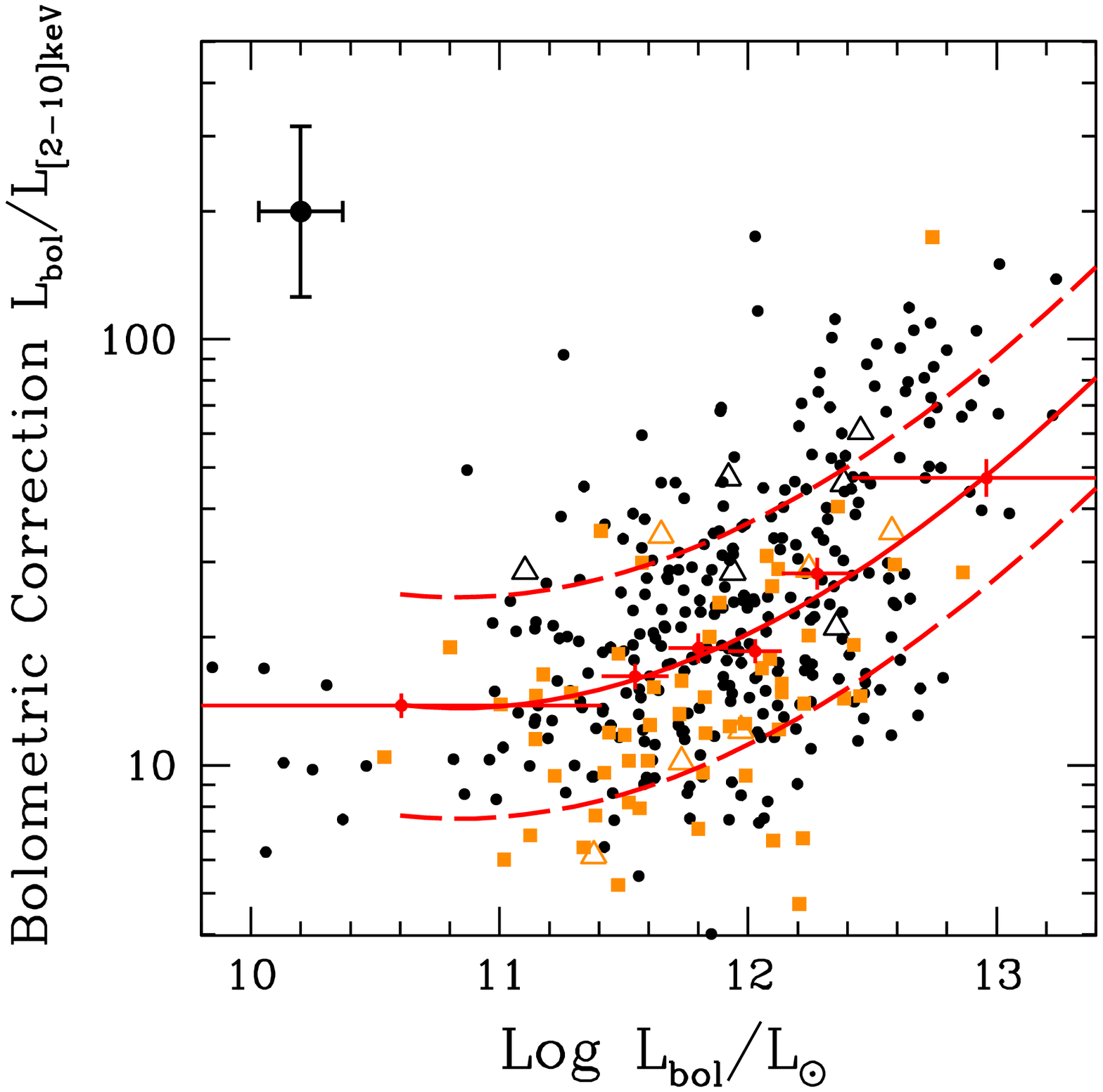}}  \\           
  {\label{fig:kbollbolbband1}\includegraphics[width=0.35\textwidth]{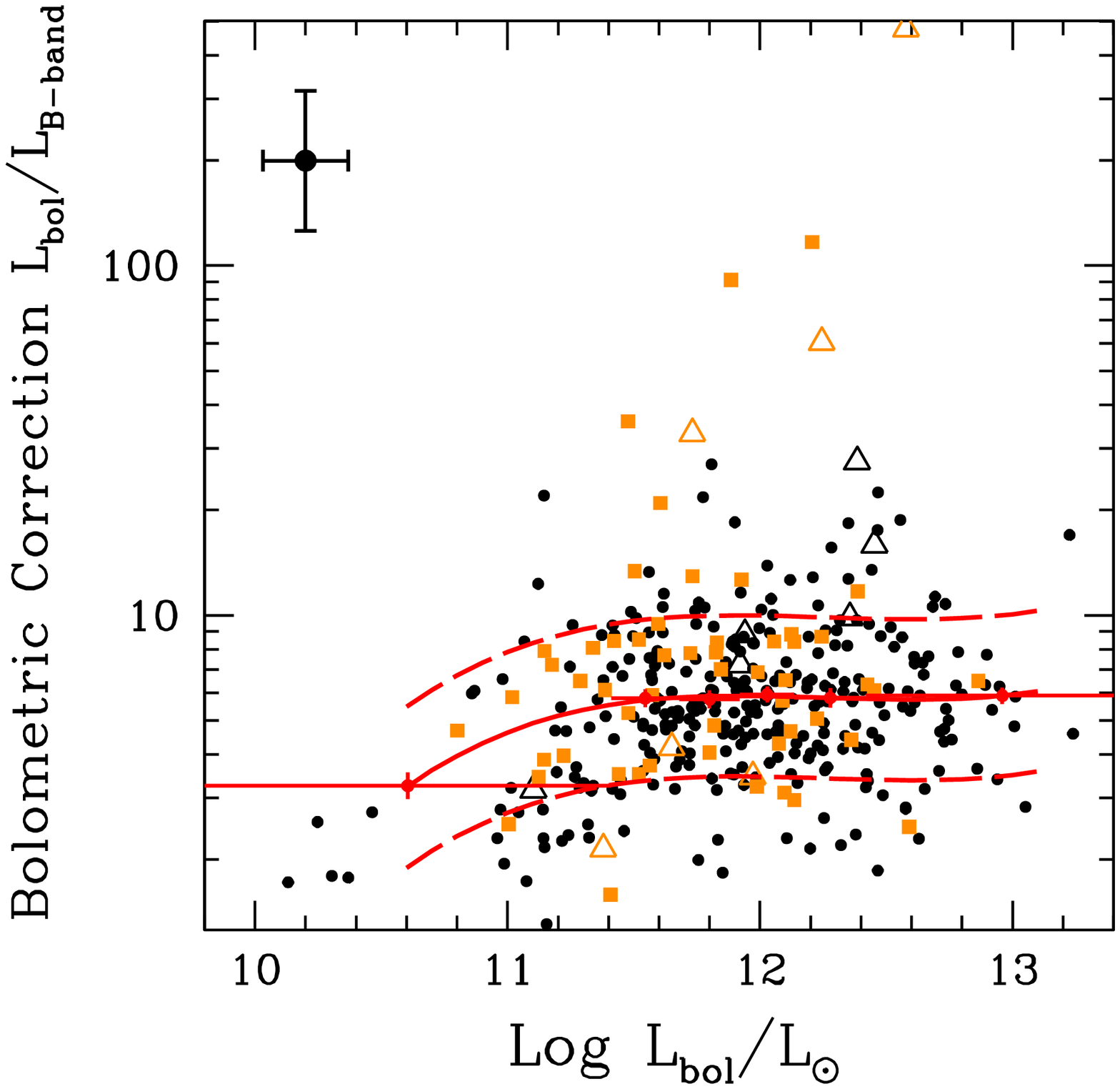}}                
  \caption{Bolometric correction as a function of the bolometric luminosity at [0.5-2]keV, [2-10]keV and in the B-band at 0.44$\mu$m for the Type-1 AGN sample. The spectro-z sample with detection in both X--ray bands is represented with the black points (N=304), while sources with spectroscopic redshift and an upper limit in the soft X--ray band are represented with black open triangles (N=6). Orange symbols represent the photo-z sample (N=63). Orange squares are photometric sources with detection in both X--ray bands (N=57), while orange open triangles are Type-1 AGN that belong to the photo-z sample with an upper limit in the soft X--ray band (N=6). Red bins are computed using 373 Type-1 AGN (6 bins with about 62 sources per bin). Red points are the median of the sources in each bin, the bars in the y axis represent the error on the median (1.4826 MAD/$\sqrt{N_{bin}}$), while the bars in the x axis are the width of the bin. The red solid line represents the best-fit relations using a third-order polynomial, while the dashed lines represent $1\sigma$ dispersion after performing a 3.5$\sigma$ clipping. Bolometric corrections have uncertainties of $\sim0.2$ dex, while $\Lbol$ uncertainties are of the order of 0.17 dex, as shown by the typical error bars in the upper left of each panel.}
  \label{fig:kbollbol1}
\end{figure*}
\subsection{Bolometric correction vs. Bolometric luminosity for the Type-1 AGN sample}
\label{Bolometric correction vs. Bolometric luminosity for the Type-1 AGN sample}
We have computed the nuclear bolometric luminosities and the bolometric corrections for the B-band, the soft and hard X--ray bands considering the Type-1 AGN sample as already described in Sect.~\ref{Bolometric luminosities and bolometric corrections}. Initially, we have computed $\kbol$ as a function of $\Lbol$ using the sample of 373 Type-1 AGN (both spectro-z and photo-z samples). Subsequently, we have divided the main sample in different sub-samples. We have considered all sources with both soft and hard X--ray detections (361 Type-1 AGN), only the spectro-z Type-1 AGN sample (310 Type-1 AGN) and the spectro-z sample but with detection in both soft and hard bands (304 Type-1 AGN). For each sub-sample we have estimated the relations between $\kbol$ and $\Lbol$ in different bands. 
We have followed a two step procedure. First, we have binned the sample in order to have approximately the same number of sources in each bin.
Then, we have computed the median in each bin and we have estimated the standard deviation in the median as $\sigma_{\rm med}=1.4826$ MAD/$\sqrt{N}$ (\citealt{1983ured.book.....H}). The MAD term is the median of absolute deviation between data and the median of data (${\rm MAD}=\langle {\rm ABS}(d-\langle d\rangle)\rangle$, where $d$ are the data).
Thereby, we have fitted the median values in the six bins using a third-order polynomial relation, and the $1\sigma$ dispersion is obtained using a $3.5\sigma$ clipping method.
In Figure~\ref{fig:kbollbol1} the bolometric corrections as a function of the bolometric luminosity in the [0.5-2]keV, [2-10]keV bands and in the B-band at 0.44$\mu$m for the Type-1 AGN sample are presented. The spectro-z sample with detection in both X--ray bands is plotted with the black points, while sources with spectroscopic redshift and an upper limit in the soft X--ray band are represented with black open triangles. Orange symbols represent the photo-z sample. Orange squares are photometric sources with detection in both X--ray bands, while orange open triangles are Type-1 AGN that belong to the photo-z sample with un an upper limit in the soft X--ray band. 
The red solid line represents one best-fit relation using the entire Type-1 AGN sample.
The best-fit relations using different sub-samples are in close agreement. However, in Table~\ref{tbl-2} the $\kbol$ as a function of $\Lbol$ relations in different bands and for different sub-samples are reported for completeness. 
These relations approximately cover two orders of magnitudes in bolometric luminosities ($11\leq\Log\Lbol {\rm [L_\odot]}\leq 13$). 
Sources plotted with open symbols in Fig.~\ref{fig:kbollbol1} should be considered upper limits, given that these $\Lbol$ are computed with an upper limit in the soft X--ray band. Consequently, these AGN are likely to move towards lower $\kbol$ and $\Lbol$ in the $\kbol-\Lbol$ plane.
The effect on $\Lbol$ cannot be very large, unless the limit on the [0.5-2] keV fluxes is so low that it implies an extremely flat spectrum and therefore a total $\Lx$ (through extrapolation) too high.
Moreover, the fact that there is no significant difference between best-fits with and without upper limits implies that neither the upper limits distribution nor the number of upper limits in each bin are affecting our results.
The $\kbol-\Lbol$ relation in the B-band seems to be nearly flat. 
Moreover, the bins at the lowest bolometric luminosity decrease with decrease $\Lbol$, but this could be due to several effects. 
First, at lower luminosities the statistics is poor and the decrease may be simply a statistical effect.
Second, luminosities in the B-band might be overestimated because of the contribution of the host-galaxy emission.
As shown in \citet{2010ApJ...724L..59H,2011ApJ...733..108H} and Elvis et al. (2012, submitted) from an analysis of Type-1 AGN in \textit{XMM}-COSMOS the host-galaxy contribution in optical/near-infrared is not negligible and may be substantial for low luminosity AGN.
Therefore, bolometric corrections at 0.44~$\mu$m for low-luminosity AGN are more affected by the galaxy emission making these values uncertain and most likely to be underestimated. 
\par
We have used optical spectra in order to have an independent estimate of the possible degree of contamination by the host-galaxy light.
We generated composite spectra for each bolometric luminosity bin by averaging all the available zCOSMOS spectra included in that bin. To create the composite, each spectrum was shifted to the rest-frame
according to its redshift and normalized in a common wavelength range, always present in the observed spectral window. The composites have been then fitted using a combination of two spectra, one representing the central active nucleus, the other one describing the host galaxy. The sets of SDSS composite spectra from \cite{richards03} were chosen as representative of the quasar emission, while a grid of 39 theoretical galaxy template spectra from \citet[hereafter BC03]{2003MNRAS.344.1000B}, spanning a wide range in age and metallicity, were used to account for the stellar component.
In the two lowest luminosity bins ($10.2\leq\Log \Lbol\leq 11.8~L_\odot$), the zCOSMOS composites can be fitted only if, along with an SDSS quasar composite, a significant host galaxy component is also included. The spectroscopic host component, fitted with a bulge-dominated BC03 template, contributes about 30\% and 20\% to the total luminosity at 4400\AA~ for the first bin ($10.2\leq\Log \Lbol\leq 11.5~L_\odot$) and the second bin ($11.5\leq\Log \Lbol\leq 11.8~L_\odot$), respectively. For both luminosity bins, the quasar template adopted is the ''dust-reddened'' one (see Richards et al. 2003), the reddest of the composite set, suggesting that, along with host galaxy contamination, a fraction of our Type-1 AGN is also experiencing a significant nuclear dust extinction (see also \citealt{2006A&A...457...79G}).
In the third bin ($11.8\leq\Log \Lbol\leq 12~L_\odot$) the average $M_{\rm B}$ is of the order of -23. This value is traditionally taken as the threshold separating the Seyfert and quasar regimes (see \citealt{2006AJ....131...84V}). 
For bins of $\Log \Lbol\geq 11.8~L_\odot$ there is no detectable host galaxy component, and the zCOSMOS average spectrum is well fitted with an SDSS quasar composites alone, although again one of the reddest composite. 
\par
Summarizing, the median bolometric correction of the two bins at bolometric luminosities less than $11.8~L_\odot$ is likely to be a lower limit. After a proper correction of the B-band luminosity, bolometric corrections should increase leading to a median value closer to that predicted by M04 and H07.
\par
\rev{We did not find any relation between the bolometric correction drawn from a given band and the corresponding luminosity. This holds for both the soft and hard X--ray bands, and for the B-band as well.
We have tested whether any relationship between $\kbol$ and luminosity exists by trying all possible permutations, but also in this case the data distribution is flat (see \citealt{vasudevanfabian07} and L11 for similar results).} 

\begin{table*}
\caption{Bolometric correction relations in different bands for the X-ray selected Type-1 and Type-2 AGN samples. \label{tbl-2}} 
\centering
\begin{tabular}{@{}l c|c c c c c c|l}
Sample & N & $a_1$ & $a_2$ & $a_3$ & $b^{\mathrm{a}}$ & $\sigma$ & N$^{\mathrm{b}}$ & Band
\\ [0.5ex]
 &  & $x$ & $x^2$ & $x^3$ &  & $3.5\sigma$ clipping &  &
\\ [0.5ex]
\hline\hline\noalign{\smallskip}
\multicolumn{9}{c}{Type-1} \\
\hline\noalign{\smallskip}
                &     & 0.239 & 0.059  & -0.009 & 1.436 & 0.26 & 1  & $[0.5-2]$keV \\[1ex]
Spectro + Photo & 373 & 0.288 & 0.111  & -0.007  & 1.308 & 0.26 & 1  & $[2-10]$keV \\[1ex]
                &     & -0.011 & -0.050 & 0.065  &0.769 & 0.23 & 6 & B-band ($0.44~\mu$m) \\[1ex]
\hline\noalign{\smallskip}
                &     & 0.248 & 0.061 & -0.041 & 1.431 & 0.26 & 1 & $[0.5-2]$keV \\[1ex]
Spectro + Photo & 361 & 0.310 & 0.114 & -0.020 & 1.296 & 0.25 & 2 & $[2-10]$keV \\[1ex]
$[0.5-2]$keV detected & & -0.026 & -0.037 & 0.075 & 0.760 & 0.22 & 5 & B-band ($0.44~\mu$m) \\[1ex]
\hline\noalign{\smallskip}
                &     & 0.250 & 0.044 & -0.023 & 1.455 & 0.26 & 1 & $[0.5-2]$keV \\[1ex]
Spectro         & 310 & 0.261 & 0.108 & 0.009 & 1.331 & 0.25 & 1 & $[2-10]$keV \\[1ex]
                &     & 0.041 & -0.065 & 0.028 & 0.763 & 0.22 & 0 & B-band ($0.44~\mu$m) \\[1ex]
\hline\noalign{\smallskip}
                &     & 0.219 & 0.068 & 0.007 & 1.444 & 0.25 & 2 & $[0.5-2]$keV \\[1ex]
Spectro         & 304 & 0.245 & 0.123  & 0.024 & 1.321 & 0.25 & 1 & $[2-10]$keV \\[1ex]
$[0.5-2]$keV detected & & 0.023 & -0.051 & 0.043 & 0.754 & 0.22 & 0 & B-band ($0.44~\mu$m) \\[1ex]
\hline\hline\noalign{\smallskip}
\multicolumn{9}{c}{Type-2} \\
\hline\noalign{\smallskip}
                                  &                       & 0.217 & 0.009 & -0.010 & 1.399 & 0.27 & 2 & $[0.5-2]$keV \\[1ex]
\raisebox{1.5ex}{Spectro + Photo} & \raisebox{1.5ex}{488} & 0.230 & 0.050 & 0.001 & 1.256 & 0.25 & 2 & $[2-10]$keV \\[1ex]
\hline\noalign{\smallskip}
Spectro + Photo       &                       & 0.208 & -0.059 & -0.038 & 1.455 & 0.28 & 0 & $[0.5-2]$keV \\[1ex]
$[0.5-2]$keV detected & \raisebox{1.5ex}{341} & 0.217 & -0.022 & -0.027 & 1.289 & 0.26 & 0 & $[2-10]$keV \\[1ex]
\hline\noalign{\smallskip}
                           &                       & 0.293 & 0.0652 & 0.0029 & 1.470 & 0.25 & 1 & $[0.5-2]$keV \\[1ex]
\raisebox{1.5ex}{Spectro}  & \raisebox{1.5ex}{248} & 0.386 & 0.071 & -0.010 & 1.395 & 0.23 & 1 & $[2-10]$keV \\[1ex]
\hline\noalign{\smallskip}
Spectro               &                       & 0.275 & 0.104 & 0.017 & 1.459 & 0.25 & 0 & $[0.5-2]$keV \\[1ex]
$[0.5-2]$keV detected & \raisebox{1.5ex}{180} & 0.411 & 0.086 & -0.010 & 1.395 & 0.23 & 0 & $[2-10]$keV \\[1ex]
\hline\hline
\end{tabular}
\flushleft\begin{list}{}
 \item[${\mathrm{a}}$] $y=a_1~x+a_2~x^2+a_3~x^3+b$, where $x=\Log L-12$ ($L$ is the bolometric luminosity in units of $L_\odot$) and $y=\Log [L/L_{\rm Band}]$.
 \item[] [${\mathrm{b}}$] Number of objects removed from the $\sigma$~clipping method.
\end{list}
\end{table*}

\begin{figure*}
  \centering
  {\label{fig:kbollbolsoft2}\includegraphics[width=0.35\textwidth]{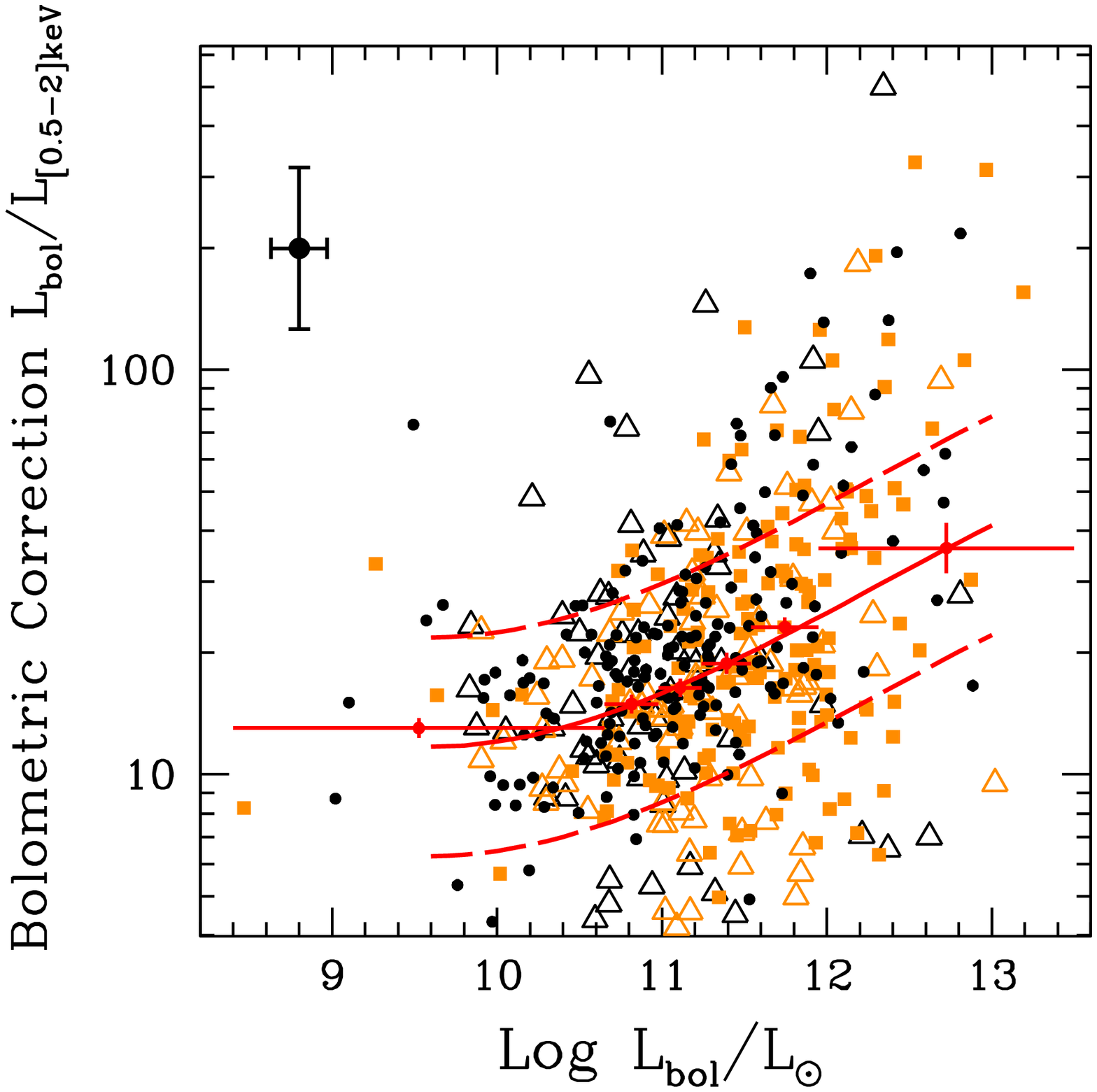}}  \qquad
  {\label{fig:kbollbolhard2}\includegraphics[width=0.35\textwidth]{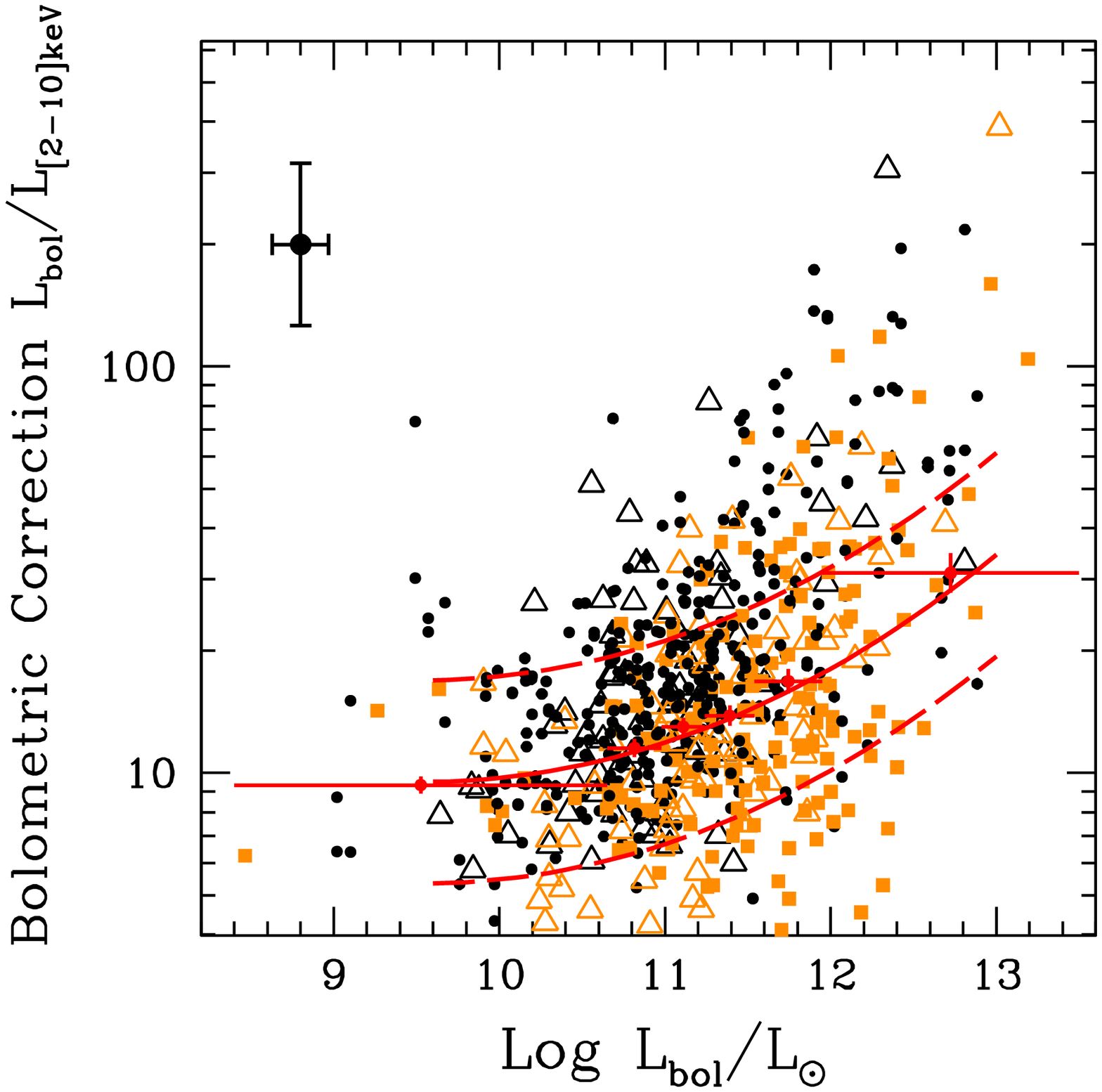}}                        
  \caption{Bolometric correction as a function of the bolometric luminosity in the [0.5-2]keV and [2-10]keV bands for the Type-2 AGN sample with AGN best-fit. Symbols key are as in Fig.~\ref{fig:kbollbol1}. The sample used to compute the bins is composed as follow: 180 spectroscopic Type-2 AGN with both X--ray detections (\textit{black points}), 68 objects with spectro-z and [0.5-2]keV upper limits (\textit{black open triangles}), 161 photo-z Type-2 AGN with both X--ray detections (\textit{orange squares}), 79 photo-z Type-2 AGN with [0.5-2]keV upper limits (\textit{orange open triangles}). Red bins are computed using the 488 Type-2 AGN sample (6 bins with about 80 sources per bin).}
  \label{fig:kbollbol2}
\end{figure*}

\subsection{Bolometric correction vs. Bolometric luminosity for the Type-2 AGN sample}
\label{Bolometric correction vs. Bolometric luminosity for the Type-2 AGN sample}
The same analysis presented in the previous Section has been applied to the Type-2 AGN sample. From the main sample of 547 Type-2 AGN, 488 AGN ($\sim89\%$) have an estimate of the bolometric luminosity, and therefore an estimate of the bolometric correction is available from the SED-fitting code.
For obscured AGN the optical emission is mostly dominated by the host-galaxy, hence we cannot estimate the nuclear luminosity in the B-band at 0.44$\mu$m.
The 488 objects have been divided in subsamples as already done for Type-1 AGN. 
The sample is composed by 180 spectroscopic Type-2 AGN with both X--ray detections, 68 objects with spectro-z and [0.5-2]keV upper limits, 161 photo-z Type-2 AGN with both X--ray detections, and 79 photo-z Type-2 AGN with [0.5-2]keV upper limits.
In Figure~\ref{fig:kbollbol2} the bolometric corrections as a function of the bolometric luminosity in the [0.5-2] and [2-10]keV bands for the Type-2 AGN sample are presented. 
These relations cover more than two orders of magnitudes ($10\leq\Log\Lbol {\rm [L_\odot]}\leq 12.5$).
Also for the Type-2 AGN sample, the best-fit relations using different sub-samples are not significantly different. In Table~\ref{tbl-2} the $\kbol$ as a function of $\Lbol$ relations in the X--ray bands and for different sub-samples are reported.

\begin{figure*}
  \centering
  {\label{}\includegraphics[width=0.35\textwidth]{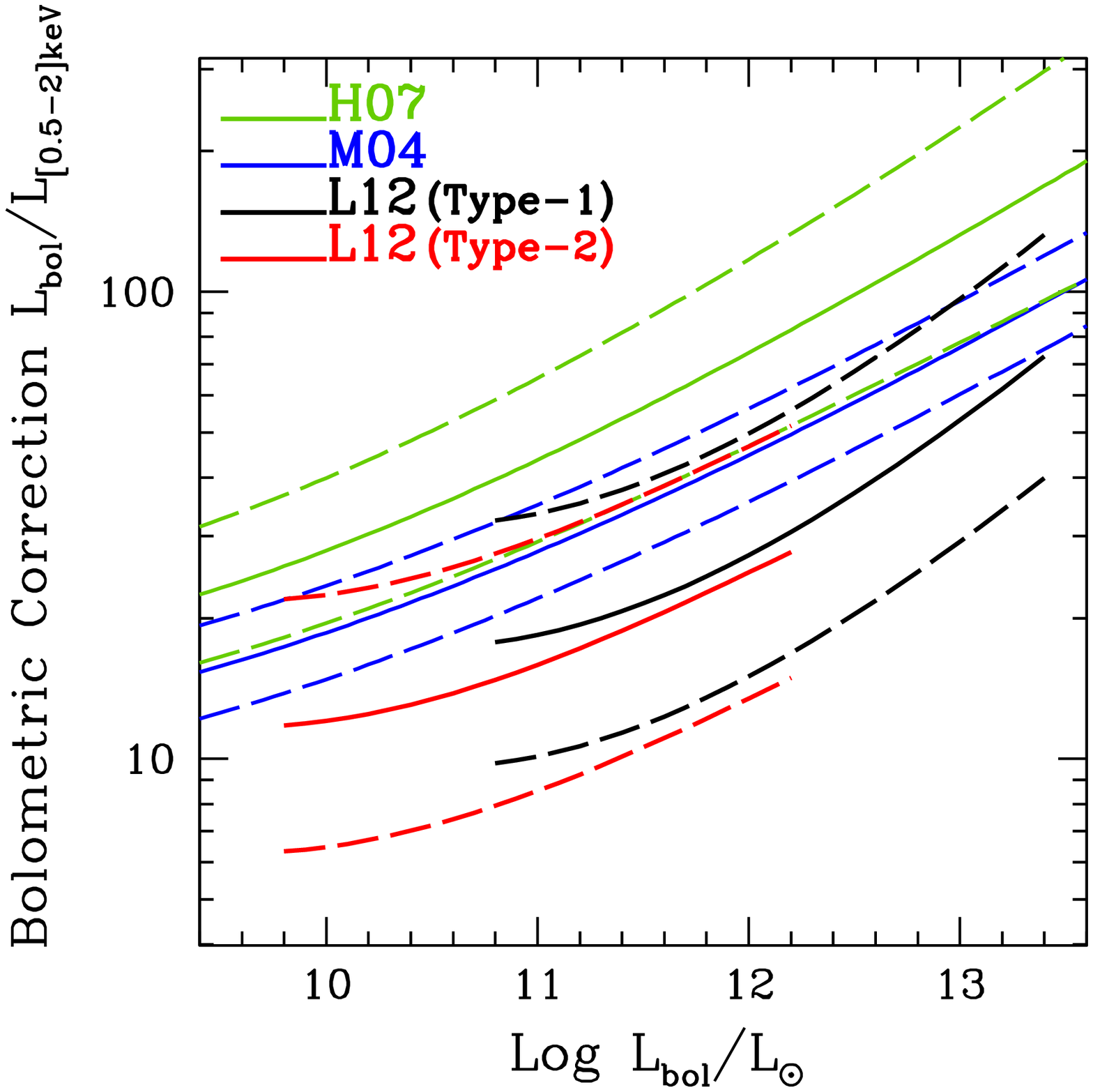}}  \qquad
  {\label{}\includegraphics[width=0.35\textwidth]{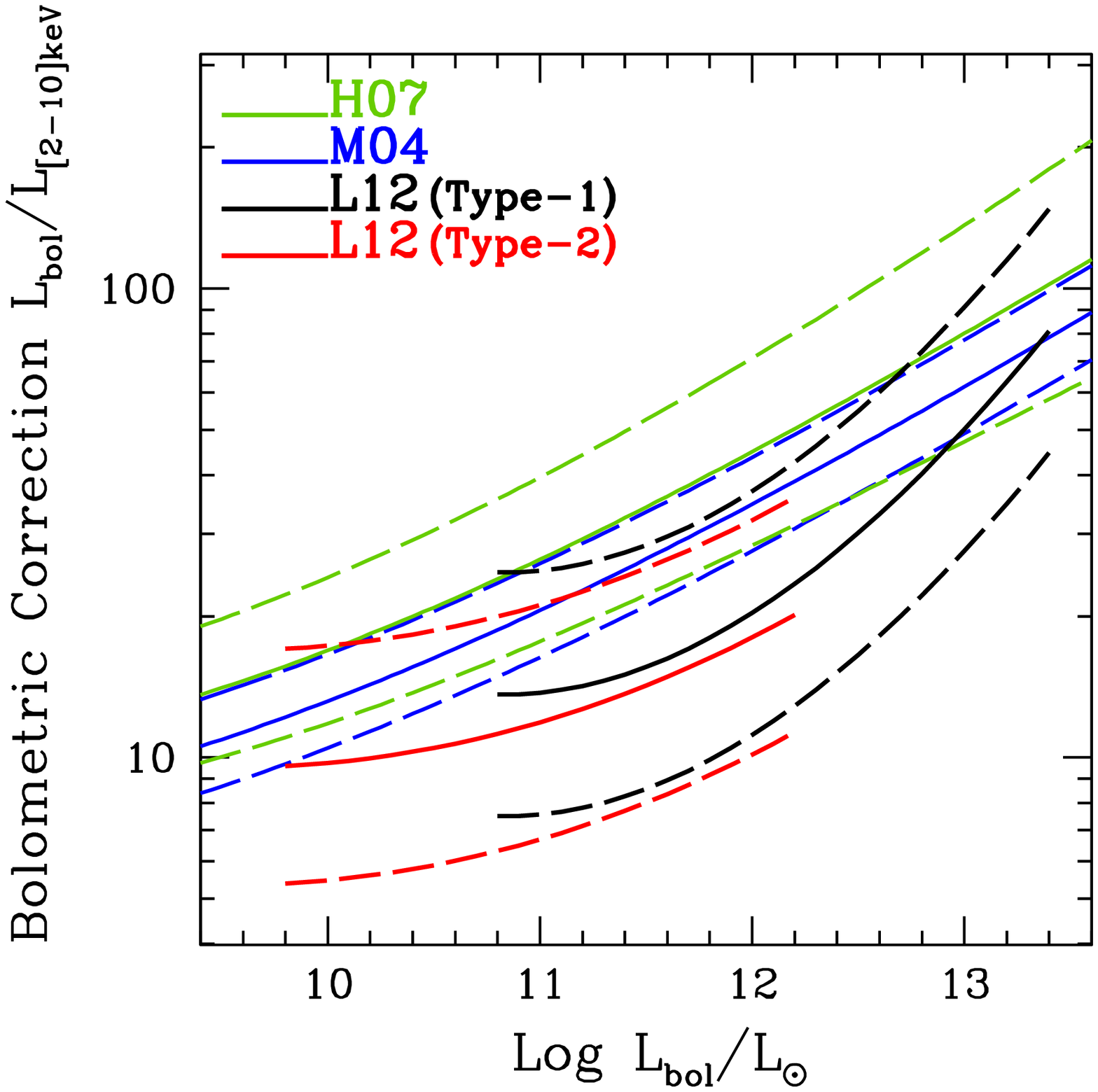}}  \\
  {\label{}\includegraphics[width=0.35\textwidth]{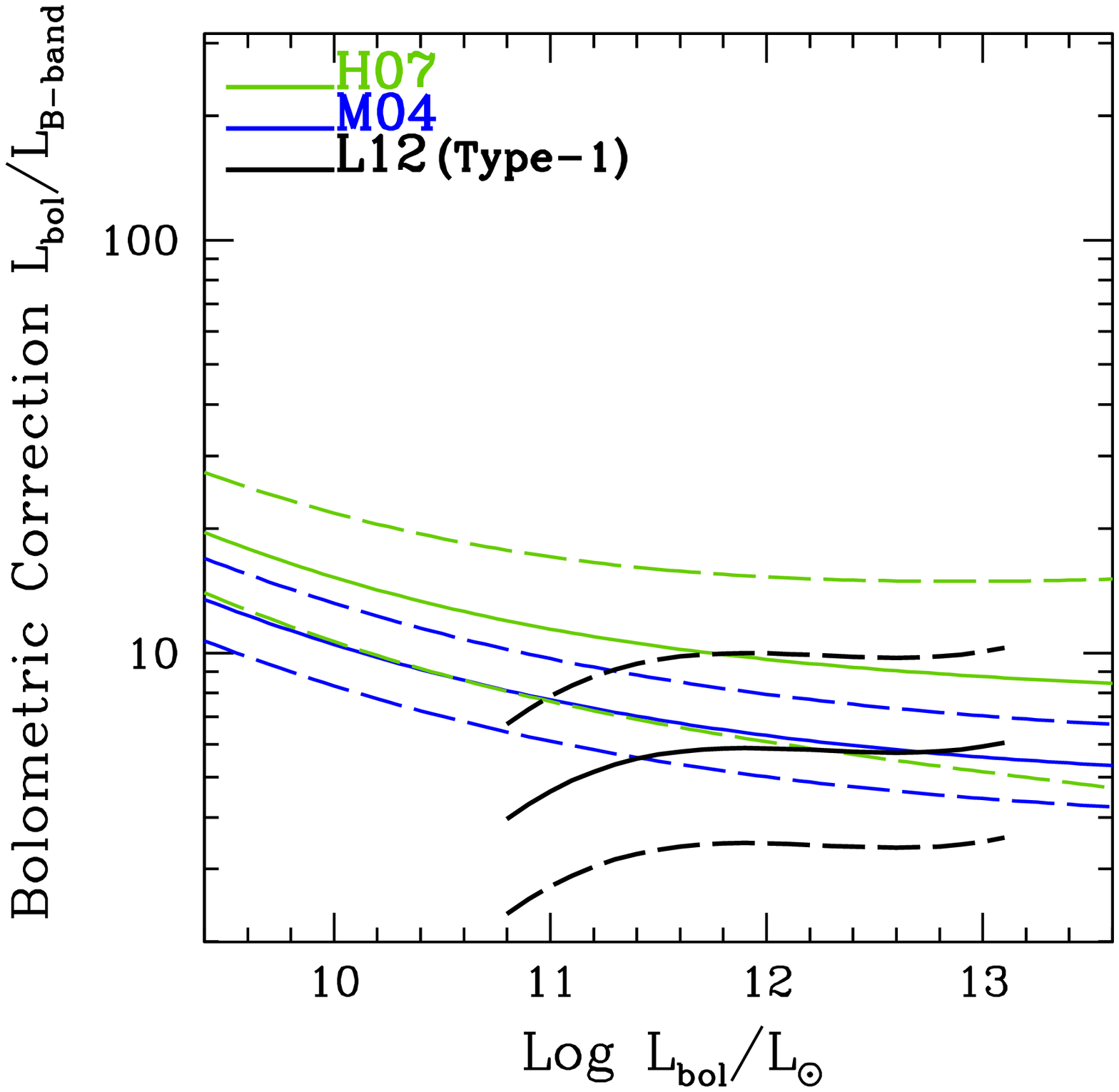}}                      
  \caption{Bolometric correction as a function of the bolometric luminosity in the soft, hard and B-bands for the Type-1 AGN sample (N=373, \textit{black solid line}), soft and hard bands for the Type-2 AGN sample with AGN best-fit (N=488, \textit{red solid line}). The $1\sigma$ dispersion after performing a 3.5$\sigma$ clipping is also reported with the dashed lines. The green and blue lines represent the bolometric correction in the hard band and $1\sigma$ dispersion obtained by H07 and M04, respectively.}
  \label{compsofthard}
\end{figure*}

\subsection{Bolometric correction vs. Bolometric luminosity: comparison of the results for Type-1 and Type-2 AGN}
\label{Bolometric correction vs. Bolometric luminosity: comparison}
Figure~\ref{compsofthard} shows the bolometric correction as a function of the bolometric luminosity in the [0.5-2]keV, [2-10]keV bands and in the B-band with $1\sigma$ dispersions after performing a 3.5$\sigma$ clipping, for the Type-1 AGN sample (N=373) and for the Type-2 AGN sample with AGN best-fit (N=488), respectively.
As a comparison, the predicted curves obtained by H07 and M04 in the soft, hard X--ray bands and in the B-band with $1\sigma$ dispersion are also reported. 
Despite the large scatter, the trend of increasing bolometric correction at increasing bolometric luminosity is confirmed. Moreover, it is evident that the $\kbol-\Lbol$ relations by M04 and H07 are higher than those observed in our (X-ray selected) samples for both Type-1 and Type-2 AGN. 
\rev{The different normalization of the $\kbol-\Lbol$ relation between H07 and M04 has to be ascribed to a different definition of bolometric luminosity. H07 include infrared wavelengths as they are interested in an empirical bolometric correction, while $\kbol$ values in M04 are estimated neglecting the optical-UV-X-ray luminosity reprocessed by the dust and therefore are representative of the AGN accretion power.
In the present analysis we have also considered only the accretion powered luminosity and thus our values should be compared with the M04 curves.
The normalization shift between our values for the $\kbol-\Lbol$ relationships and the M04 curves could be, at least partly, due to a selection effect. Since our sample is hard X-ray selected, it is biased towards X-ray bright objects with lower bolometric corrections. The SED adopted by M04 is typical of luminous optically selected quasars, therefore it is biased towards higher $\kbol$ values.}
\par
The present results may have interesting consequences. In fact all accretion models, that also include mergers, fail in reproducing the high mass end of the local black hole mass function (see \citealt{2009ApJ...690...20S}). However, as suggested by our data, assuming a lower $\kbol$ than the one inferred by M04 or H07, can ease the tension between models and data (see discussion in \citealt{2011arXiv1111.3574S}).
The range of validity of these curves are limited to slightly more than two orders of magnitudes for both Type-1 and Type-2 AGN, where $\Lbol$ ranges from $10^{10}L_\odot$ to $10^{12}L_\odot$ for Type-2 AGN, and from $10^{11}L_\odot$ to $10^{13}L_\odot$ for Type-1 AGN.
In the overlapping range of bolometric luminosity ($10^{11-12}L_\odot$), there is no significant difference between the bolometric corrections for Type-1 and Type-2 AGN.
It is also interesting and noteworthy that the $\kbol-\Lbol$ relations for Type-2 AGN seem to be the natural extension of the Type-1 relations at lower luminosities. 
Even if we can explore a limited range of bolometric luminosities in both AGN samples this relationship can be applied for all AGN across nearly four decades in luminosity.
\par
As a final comment, we want to discuss a comparison between the results presented in this paper and the results on the $\alphaox$ parameter in L10.
\rev{The fact that $\alphaox$ and $\kbol$ show a similar behaviour is a natural consequence of the tight correlation between these two parameters, which has been discussed in depth in L10. 
Indeed, $\alphaox$ is almost independent on the X--ray luminosity at 2 keV, while there is a strong trend with the optical luminosity at 2500\AA{} (see also \citealt{avnitananbaum79,zamorani81,vignali03,steffen06,just07,2009ApJS..183...17Y,2012A&A...539A..48M}), which is a tracer of the bolometric luminosity (more than 70\% of $\Lbol$ comes from the optical-UV). It is not surprising then to find a correlation between $\kbol$ (evaluated in the soft and the hard X--ray bands) and $\Lbol$, while no correlation is seen with the X--ray luminosity. 
In conclusion, this analysis suggests that the fundamental underlying correlation arises between $\kbol$ and $\Lbol$.}

\begin{figure}
 \includegraphics[width=8cm]{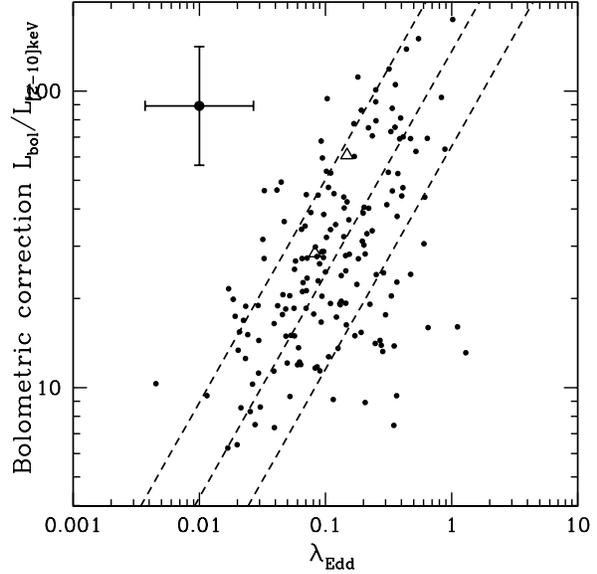}
 \caption{Hard X-ray bolometric correction versus Eddington ratio for the 170 Type-1 AGN  (\textit{black points})  with BH mass estimate from broad lines. The short-dashed black line shows the best-fit relation and 1$\sigma$ dispersion that we found using the OLS bisector algorithm. Typical error bars in the upper left of the panel are showed for both $\lambdaEdd$ (0.43 dex) and $\kbol$ (0.2 dex).}
 \label{kbolledd1}
\end{figure}
\begin{figure}
 \includegraphics[width=8cm]{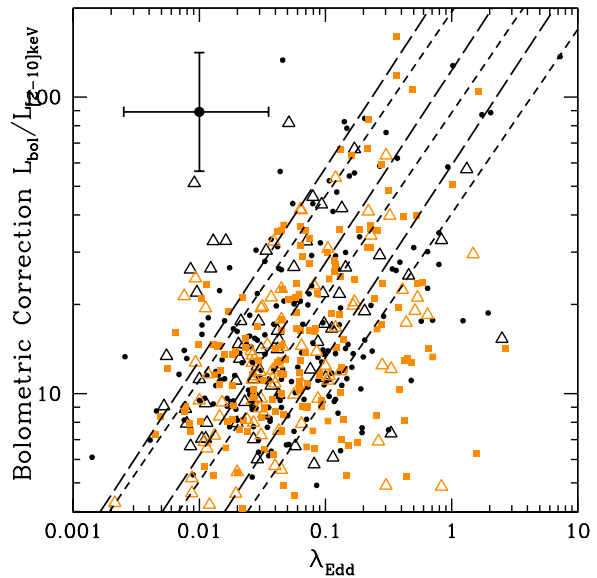}
 \caption{Hard X-ray bolometric correction versus Eddington ratio for the 488 Type-2 AGN with bolometric luminosities and stellar masses available from the SED-fitting. Symbols key are as in Fig.~\ref{fig:kbollbol1}. The short-dashed black line shows the OLS bisector and 1$\sigma$ dispersion for the 488 Type-2 AGN with $\Lbol$ and $M_\ast$ available. The long-dashed black line shows the OLS bisector and 1$\sigma$ dispersion for the 144 Type-2 AGN with $\Lbol$, $M_\ast$ and morphology classification available. Typical error bars in the upper left of the panel are showed for both $\lambdaEdd$ (0.55 dex) and $\kbol$ (0.2 dex).}
 \label{kbolledd2}
\end{figure}
\begin{figure}
 \includegraphics[width=8cm]{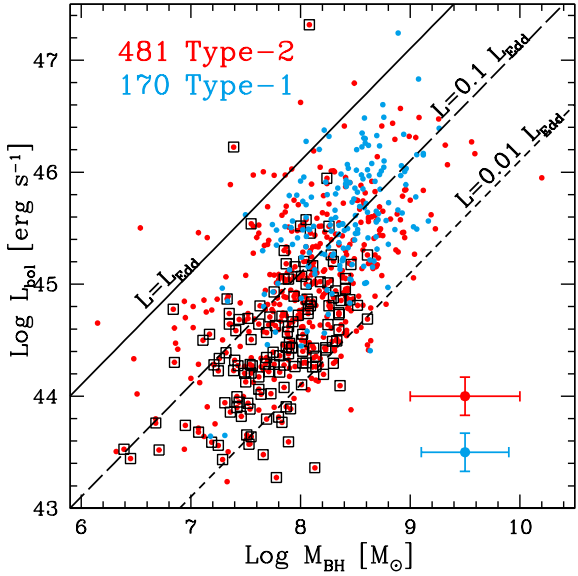}
 \caption{Bolometric luminosities as a function of black hole masses for both Type-1 (\textit{blue symbols}) and Type-2 (\textit{red symbols}) AGN samples. Diagonal lines represent the trend between $\Lbol$ and $\Mbh$ at different fractions of the Eddington luminosity: $\lambdaEdd=1, 0.1$ and 0.01 (solid, long-dashed and short dashed lines, respectively). Black open squares mark 144 Type-2 AGN with $\Lbol$, $M_\ast$ and morphology classification available. Typical error bars in the lower right of the panel are showed for both $\Mbh$ (0.4 dex for Type-1 AGN in blue and 0.5 dec for Type-2 AGN in red) and $\kbol$ (0.2 dex for both Type-1 and Type-2 AGN).}
 \label{lbolmbh12}
\end{figure}

\subsection{Hard X--ray bolometric correction vs. Eddington ratio}
\label{Bolometric correction vs. Eddington ratio}
Several works in the literature found a trend between the hard X--ray $\kbol$ and $\lambdaEdd$, although with the presence of a large scatter (e.g., \citealt{vasudevanfabian07}; V09a; V09b; \citealt{2010MNRAS.402.1081V}; L10), or between $\lambdaEdd$ and the intrinsic bolometric AGN luminosity (e.g., \citealt{2009ApJ...700...49T,2011ApJ...733...60T}). The scatter is not reduced even considering AGN with simultaneous optical, UV and X-ray data retrieved from the XMM-\textit{Newton} EPIC-pn and Optical Monitor (OM) archives (see Fig.~11 in L10, \citealt{2009MNRAS.392.1124V}).
In Figure~\ref{kbolledd1} the hard X--ray $\kbol$ as a function of the $\lambdaEdd$ for Type-1 AGN is presented. 
We have computed the ordinary least-squares (OLS) bisector for the $\kbol-\lambdaEdd$ relation considering the 170 Type-1 AGN with $\Mbh$ mass estimates from broad lines (there are only two objects with an upper limit in the soft band). 
The best-fit parameters for the $\kbol-\lambdaEdd$ relation using OLS(Y$|$X) (i.e. treating $\lambdaEdd$ as the independent variable), OLS(X$|$Y) (i.e. treating the hard X--ray $\kbol$ as the independent variable) and the OLS bisector are reported in Table~\ref{tbl-3}.
We find that the slope of the bisector relation considering the 170 Type-1 AGN sample is statistically consistent with the slope found considering the subsample of 150 Type-1 AGN in L10.
\begin{table}
\caption{Hard X--ray bolometric correction as a function of the Eddington ratio for the X-ray selected 170 Type-1 AGN sample with $\Mbh$ available. \label{tbl-3}} 
\centering
\begin{tabular}{@{}c c c c |l}
 $(m\pm dm)^{\mathrm{a}}$ & $(q\pm dq)^{\mathrm{a}}$ & $\sigma$ [$3.5\sigma$ clipping] &  Relations\\ [0.5ex]
\hline\noalign{\smallskip}
$0.752\pm0.035$ & $2.134\pm0.039$ & 0.32  & OLS bisector \\[1ex]
\hline\noalign{\smallskip}
$0.392\pm0.055$ & $1.799\pm0.063$ & 0.27  & OLS(Y$|$X) \\[1ex]

\hline\noalign{\smallskip}
$1.303\pm0.149$ & $2.647\pm0.128$ & 0.49  & OLS(X$|$Y) \\[1ex]
\hline\hline
\end{tabular}

\flushleft\begin{list}{}
 \item[${\mathrm{a}}$]{ $y=(m\pm dm)~x+(q\pm dq)$, where $x=\Log \lambdaEdd$ and $y=\Log [L/L_{[2-10]{\rm keV}}]$.}
 \item[] No object has been removed from the $\sigma$~clipping method.
\end{list}
\end{table}
\par
The same analysis has been performed using 488 Type-2 AGN for which bolometric luminosities and stellar masses are available from the SED-fitting, and for the subsample of 144 Type-2 AGN with reliable morphology classification. The data are shown in Figure~\ref{kbolledd2}.
The trend of increasing Eddington ratios at increasing bolometric corrections, as found for Type-1 AGN, is confirmed also for Type-2 AGN. The slope of the bisector relation considering the 170 Type-1 AGN sample is marginally consistent at the 3$\sigma$ level with the slope found considering the 488 Type-2 AGN sample, while it is fully consistent with the slope found considering the 144 Type-2 AGN sample \rev{with reliable morphological classifications}. Also the normalizations of the $\kbol-\lambdaEdd$ relations for the Type 1 and Type 2 AGN are in good agreement with each other. The best-fit parameters for the $\kbol-\lambdaEdd$ relation using OLS(Y$|$X), OLS(X$|$Y) and the OLS bisector are reported in Table~\ref{tbl-4}.
\par 
The hard X-ray radiation of AGN with relatively high Eddington ratio is commonly thought to be produced from a disk corona as a result of Comptonization of soft photons arising from the accretion disk (e.g. \citealt{1991ApJ...380L..51H,1993ApJ...413..507H,2001ApJ...546..966K,2009MNRAS.394..207C}). If the bolometric luminosity is the result of accretion disk and corona emission, the fraction of X--ray luminosity over the total luminosity represents the strength of the corona relative to the accretion disk. 
Therefore, the correlation between the hard X-ray bolometric correction and the Eddington ratio for Type-1 and Type-2 AGN indicates that the corona relative to the disk becomes weaker as the Eddington-scaled accretion rate increases.
\par
In Figure~\ref{lbolmbh12} the bolometric luminosities are plotted as a function of black hole masses for both Type-1 and Type-2 AGN samples. In the following we will focus on the sources which are undoubtedly dominated by AGN activity (i.e. we have removed 7 sources from the Type-2 AGN sample with $\Log \Lhard<42$ erg s$^{-1}$). Diagonal lines represent the trend between $\Lbol$ and $\Mbh$ at different fractions of the Eddington luminosity ($\lambdaEdd=1, 0.1$ and 0.01).
It is worth noting that black hole masses for Type-1 and Type-2 AGN are derived in completely different ways (i.e., virial estimators versus scaling relations), and the difference between black hole masses (and Eddington ratios) for Type-1 and Type-2 AGN would decrease if a Salpeter IMF were used to compute stellar masses for Type-2 AGN. Stellar masses computed with the Salpeter IMF would increase by a factor $\sim$1.7. As a consequence, black hole mass and Eddington ratio estimates would increase by a similar factor.  
There is a continuity between bolometric luminosities as a function of $\Mbh$ for Type-1 and Type-2 AGN, where few sources have $\Lbol$ lower than 0.01 $\Ledd$. 
To determine whether Eddington ratios are affected by any significant evolution, we have studied a possible dependence of $\lambdaEdd$ on redshift, $\Lbol$, X--ray luminosities, $\Mbh$ and column densities. We find no correlation between Eddington ratios with both X--ray luminosities and column densities. The other correlations are discussed in the following.

\begin{table}
\caption{Hard X--ray bolometric correction as a function of the Eddington ratio for the X-ray selected Type-2 AGN sample. \label{tbl-4}} 
\centering
\begin{tabular}{@{}c c c c |l}
 $(m\pm dm)^{\mathrm{a}}$ & $(q\pm dq)^{\mathrm{a}}$ & $\sigma$ [$3.5\sigma$ clipping] & N$^{\mathrm{b}}$ & Relations\\ [0.5ex]
\hline\hline\noalign{\smallskip}
\multicolumn{5}{c}{488 Type-2$^{\mathrm{c}}$} \\
\hline\noalign{\smallskip}
$0.621\pm0.025$ & $1.947\pm0.035$ & 0.34 & 2 & OLS bisector \\[1ex]
\hline\noalign{\smallskip}
$0.283\pm0.036$ & $1.536\pm0.052$ & 0.26 & 4 & OLS(Y$|$X) \\[1ex]
\hline\noalign{\smallskip}
$1.105\pm0.084$ & $2.536\pm0.104$ & 0.57 & 0 & OLS(X$|$Y) \\[1ex]
\hline\hline\noalign{\smallskip}
\multicolumn{5}{c}{144 Type-2$^{\mathrm{d}}$} \\
\hline\noalign{\smallskip}
$0.649\pm0.054$ & $2.090\pm0.084$ & 0.32 & 0 & OLS bisector \\[1ex]
\hline\noalign{\smallskip}
$0.375\pm0.097$ & $1.702\pm0.151$ & 0.23 & 3 & OLS(Y$|$X) \\[1ex]
\hline\noalign{\smallskip}
$1.015\pm0.095$ & $2.608\pm0.137$ & 0.44 & 1 & OLS(X$|$Y) \\[1ex]
\hline\hline
\end{tabular}

\flushleft\begin{list}{}
 \item[${\mathrm{a}}$]{ $y=(m\pm dm)~x+(q\pm dq)$, where $x=\Log \lambdaEdd$ and $y=\Log [L/L_{[2-10]{\rm keV}}]$.}
 \item[] [${\mathrm{b}}$] Number of objects removed from the $\sigma$~clipping method.
 \item[] [${\mathrm{c}}$] Type-2 AGN with $\Lbol$ and $M_\ast$ available.
 \item[] [${\mathrm{d}}$] Type-2 AGN with $\Lbol$, $M_\ast$ and morphology classification available.
\end{list}
\end{table}

\subsection{Luminosity-redshift dependence of Eddington ratio}
\label{Luminosity-redshift dependence}
\begin{figure}
 \includegraphics[width=8cm]{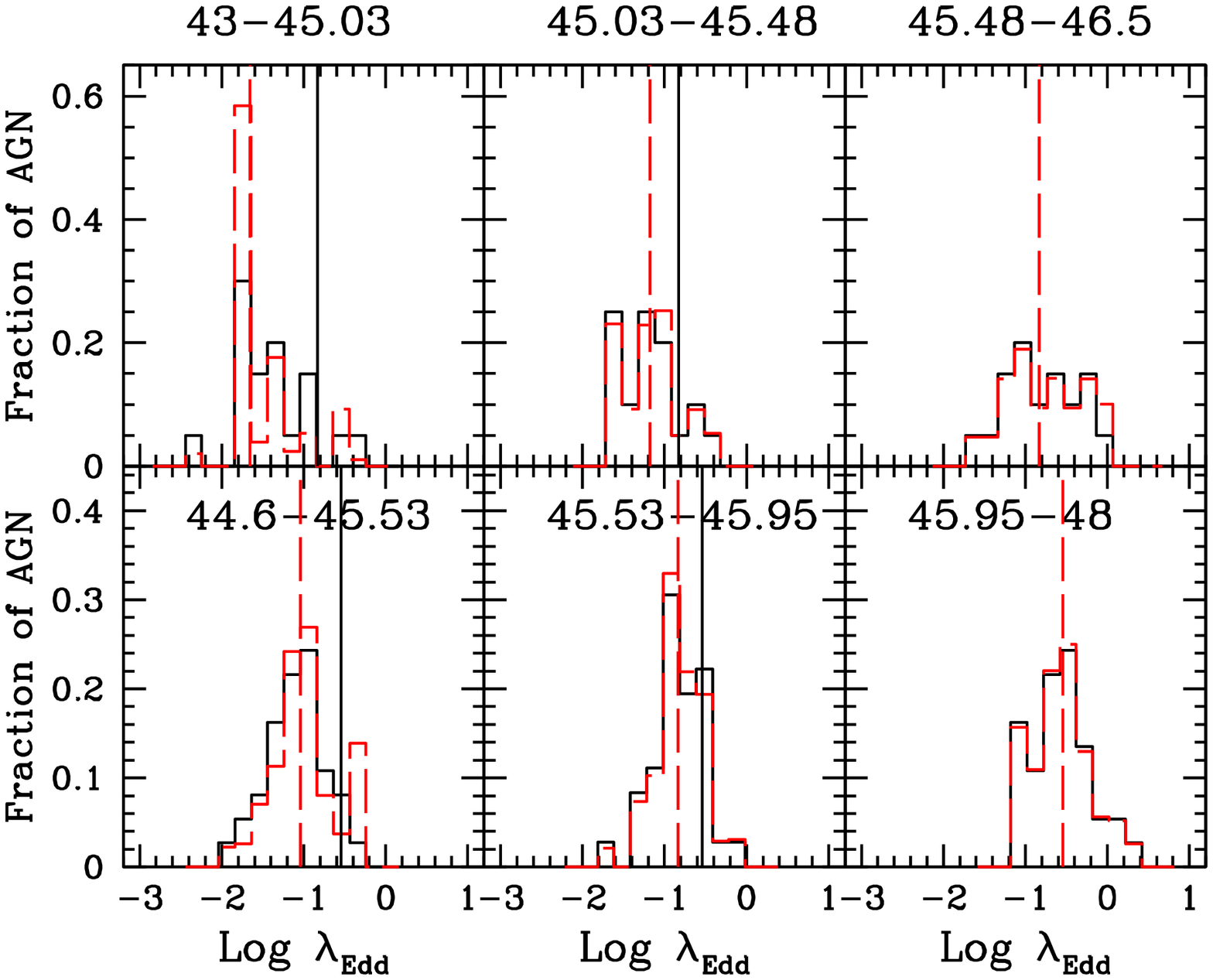}
 \caption{Distributions of Eddington ratios in bins of luminosity and redshift for the Type-1 AGN sample. The panels are divided between $z<1.2$ (top three panels) and $1.2\leq z \leq 2.3$ (lower three panels), and bolometric luminosities increases from left to right ($\Log\Lbol$ [erg s$^{-1}$] intervals are reported on top of each panel). The black histograms show the observed $\lambdaEdd$ distributions, while the red dashed histograms represent the distributions corrected for incompleteness, according to the $V/V_{\rm max}$ method described in $\S$~\ref{Luminosity-redshift dependence}. The dashed red lines are the median values corresponding to the completeness-corrected distributions, while the solid black lines in the first and in the second bins are plotted at the $\lambdaEdd$ value corresponding to the median in the highest luminosity bin.  There are $\sim20$ and $\sim37$ objects in each bin at $z<1.2$ and $1.2\leq z \leq 2.3$, respectively. The numbers of AGN in the completeness-corrected $\lambdaEdd$ distributions are 98, 22, and 21 at $z<1.2$, while there are 88, 51, and 38 at $1.2\leq z \leq 2.3$, from left to right.}
 \label{panelleddbin1}
\end{figure}

\begin{figure}
 \includegraphics[width=8cm]{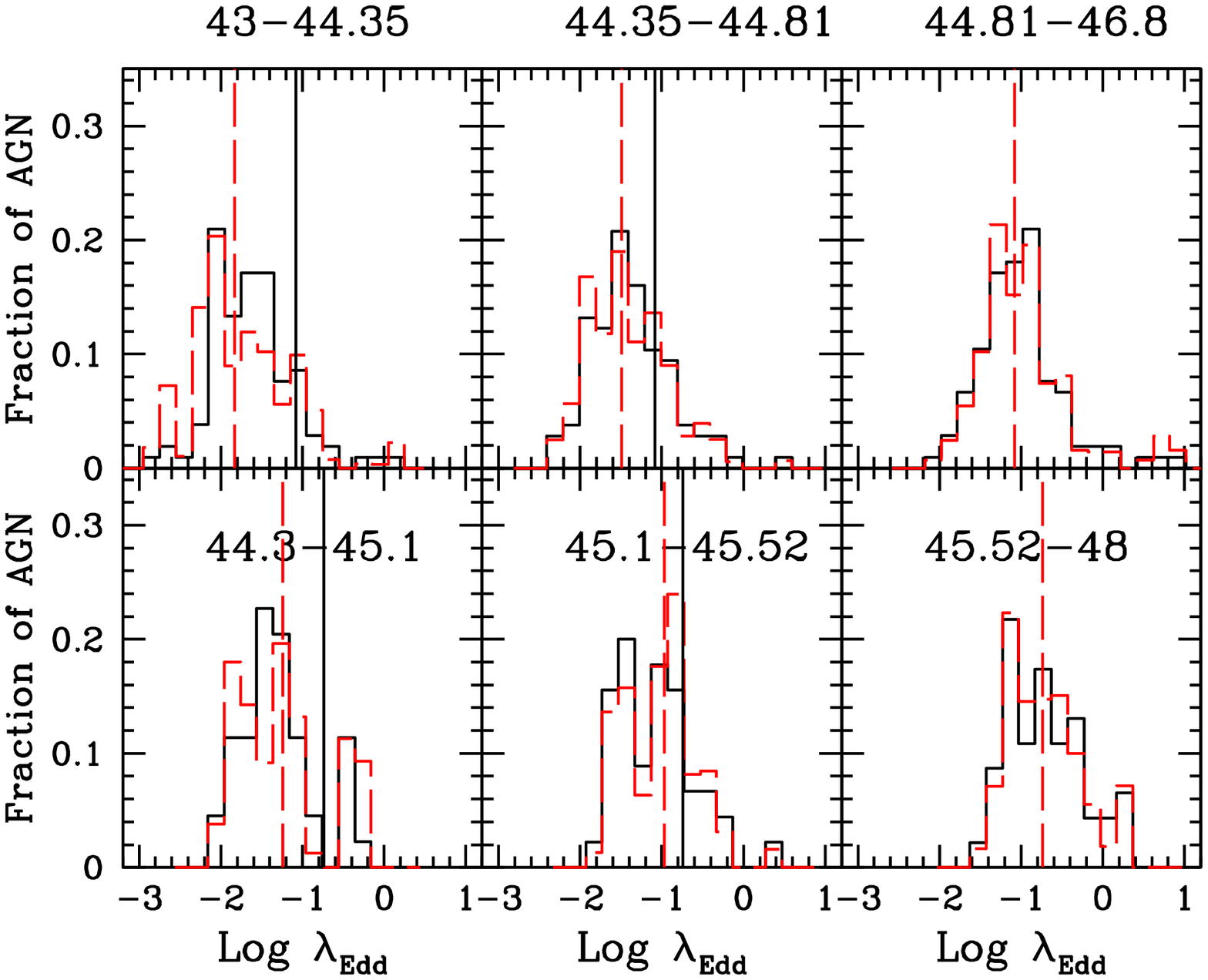}
 \caption{Distributions of Eddington ratios in bins of luminosity and redshift for the Type-2 AGN sample. Description as in Fig.~\ref{panelleddbin1}. There are $\sim106$ and $\sim45$ objects in each bin at $z<1.2$ and $1.2\leq z \leq 2.3$, respectively. The numbers of AGN in the completeness-corrected $\lambdaEdd$ distributions are 1317, 234, and 140 at $z<1.2$, while there are 259, 77, and 62 at $1.2\leq z \leq 2.3$ from left to right.}
 \label{panelleddbin2}
\end{figure}

We have explored the possibility of a dependency of $\lambdaEdd$ with redshift and bolometric luminosity by binning both Type-1 and Type-2 AGN samples in $z$ and $\Lbol$. The samples are divided in two redshift bins and three $\Lbol$ bins. The redshift bins are $z<1.2$ and $1.2\leq z \leq 2.3$, while the luminosity cuts for each sample are chosen in order to have approximately the same number of objects in each bin. The redshift bins have been defined in order to sample the observed evolution of the hard X-ray luminosity function of AGN determined by \citet[see their Fig.~9 and the discussion below]{2010MNRAS.401.2531A}. There are 60 Type-1 AGN and 317 Type-2 AGN at $z<1.2$, while 109 Type-1 AGN and 135 Type-2 AGN are at $1.2\leq z \leq 2.3$. At $z>2.3$ the number of AGN is not large enough to be statistically significant.
The black histograms in Figs.~\ref{panelleddbin1} and \ref{panelleddbin2} show the observed $\lambdaEdd$ distributions for Type-1 and Type-2 AGN, respectively.
\par
\rev{The observed $\lambdaEdd$ distributions may be biased by selection effects related to the depth of the X--ray data, and the fall-off below the peak at  low $\lambdaEdd$ can be partly due to incompleteness of the X--ray selection. We have therefore quantified the impact of this incompleteness on our $\lambdaEdd$ distribution for each bin by employing the standard $V_{\rm max}$ method, introduced by \citet{1968ApJ...151..393S}. The quantity $V_{\rm max}$ represents the maximum volume where an object would still be detectable in our survey given its X--ray luminosity, redshift and column density and is described by 
\begin{equation}
V_{\rm max}(i)=\int_{z_{\rm min}(i)}^{z_{\rm max}(i)} \Omega(L_{\rm X}(i),z(i),\NH(i)) (1+z)^k\frac{{\rm d}V}{{\rm d}z} {\rm d} z,
\end{equation}
where $z_{\rm min}$ is the lower boundary of the redshift bin, $z_{\rm max}$ is the minimum between the upper boundary of the redshift bin and the redshift where the $i^{\rm th}$ object would not be detectable anymore in the survey. The parameter $\Omega$ is the solid angle covered by XMM-\textit{Newton} at the flux level $f(L_{\rm X}(i),z)$, and ${\rm d}V/{\rm d}z$ is the comoving volume.
The term $(1+z)^k$ describes the AGN evolution, which we have chosen to represent as a pure density evolution\footnote{The results would be similar if we adopted a different evolutionary law.}.
At zeroth-order, we have neglected any $\NH$ dependence and, therefore, we have estimated the flux at each redshift considering a simple k-correction
\begin{equation}
f(L_{\rm X}(i),z)=\frac{L_{\rm X}(i) (1+z)^{(1-\alpha)}}{4\pi {\rm d_L^2}},
\end{equation}
where $\alpha=0.7$ \citep{2005ApJ...635..864L} and $L_{\rm X}$ is the de-absorbed X--ray luminosity in the $[2-10]$ keV rest-frame band.
To obtain the area at each X--ray flux we have employed the $[2-10]$ keV sky coverage computed by \citet[see their Fig.~5]{cappelluti09}, and we have finally integrated over the comoving volume. 
The $k$ value in the evolutionary term has been chosen to match the observed evolution of the hard X-ray luminosity function of AGN determined by \citet[see their Fig.~9]{2010MNRAS.401.2531A}. In the low redshift bin we have considered $k=(2.2, 2.4, 4.3, 7.8)$ for $\Log \Lhard=(42, 43, 43.5, 44)$ erg s$^{-1}$, while for $\Log \Lhard>44$ erg s$^{-1}$ we have used 7.8\footnote{The choice of this value is not critical because all objects at $\Log \Lhard>44$ erg s$^{-1}$ have $V/V_{\rm max}\simeq1$.}. 
We have adopted $k=0$ consistently with the luminosity function at $z>1.2$ by  Aird et al. for the range of luminosities covered by our sample.
For each AGN we have adopted the appropriate $k$ value interpolating the numbers above.
The total volume ($V$) has been estimated through the same procedure, where now the area is the total area covered by XMM-\textit{Newton} (i.e. 2.13 deg$^2$).
The $\lambdaEdd$ distributions weighted by the ratio $V/V_{\rm max}$ for each object are plotted with the red dashed histograms in Figs.~\ref{panelleddbin1} and \ref{panelleddbin2}. As shown in the figures, the completeness correction does not change significantly the histograms.}
There are several interesting points to note.
\par
First, the distribution of $\Log \lambdaEdd$ is nearly Gaussian, especially at high redshift and high luminosity, with a dispersion of the order of $\sim0.35$ dex for the Type-1 AGN sample and $\sim0.5$ dex for the Type-2 AGN sample. 
As expected, the low redshift/luminosity bins are more sensitive to incompleteness.
\par
Second, it is evident that the population of AGN that we are studying is dominated by sub-Eddington accretion rate objects.
This result in contrast with the findings obtained by \citet{2006ApJ...648..128K}, where the AGN population is dominated by near-Eddington accretors. 
The difference, consistently with the trend that we see in our data (see below), might be due to the fact that the bulk of the AGN population studied by Kollmeier and collaborators have higher $\Lbol$, typically in the range $\Lbol\sim10^{45-47}$ erg s$^{-1}$. 
\par
\begin{figure}
 \includegraphics[width=8cm]{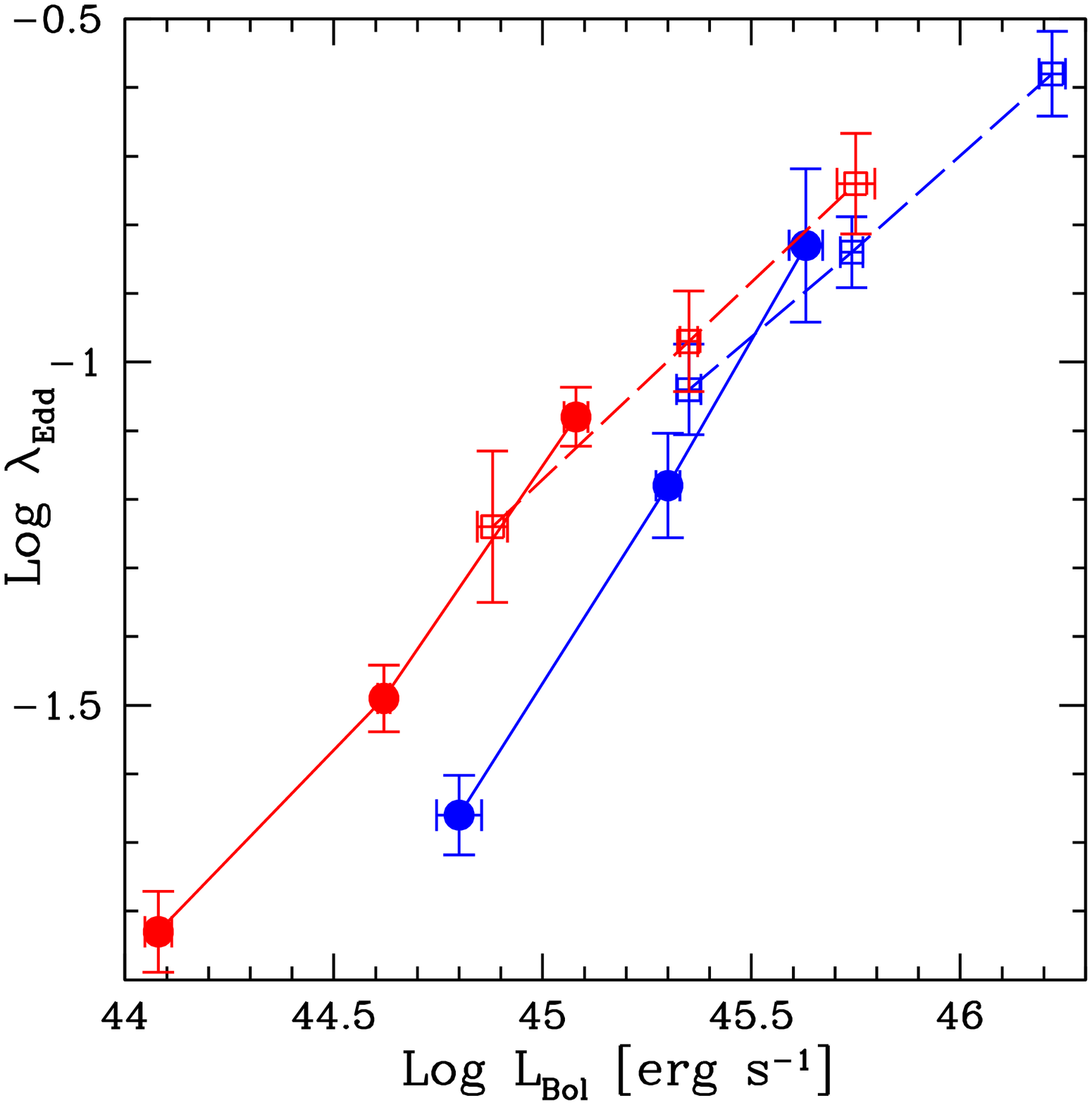}
 \caption{Median Eddington ratios as a function of the median $\Lbol$ for Type-1 (\textit{blue symbols}) and Type-2 AGN (\textit{red symbols}). Filled circles and open squares represent the median $\lambdaEdd$ for $z<1.2$ and $1.2\leq z \leq 2.3$, respectively. Solid lines connect low redshift bins, while dashed ones connect high redshift bins. Error bars on the median are estimated considering the 16th and 84th percentile divided by the square root of the number of observed AGN.}
 \label{lbolledd}
\end{figure}

Third, the Eddington ratio increases with luminosity for both Type-1 and Type-2 AGN. 
In Fig.~\ref{lbolledd} the median $\lambdaEdd$ is plotted against the median $\Lbol$ for both AGN populations. Different symbols for low redshift (filled circles) and high redshift (open squares) are introduced. 
There is no clear evolution of $\lambdaEdd$ with redshift in both redshift bins.
At a given $\Lbol$, Type-2 AGN seem to have higher $\lambdaEdd$ than Type-1 AGN at low redshift, while at high redshift the difference is not significant. 
A summary of the average $\Log\lambdaEdd$ values and relative dispersions for the Type-1 and Type-2 AGN samples are given in Tables~\ref{tbl-4} and \ref{tbl-5}, respectively.
\par
Summarizing, we have found that Eddington ratio evolves with bolometric luminosity for both Type-1 and Type-2 AGN, while it does not show a clear evolution in redshift if we bin in $\Lbol$. Type-1 AGN have median Eddington ratios ranging, on average, from $\Log \lambdaEdd\sim-1.6$ to $\Log \lambdaEdd\sim-0.6$ across the luminosity scale (with a dispersion of $\sim0.35$ dex), while the corresponding values for Type-2 AGN range from $\Log \lambdaEdd\sim-1.8$ to $\Log \lambdaEdd\sim-0.7$ (with a dispersion of $\sim0.5$ dex). 

\begin{table}
\caption{Average $\Log\lambdaEdd$ values in $\Lbol-z$ bins for Type-1 AGN. \label{tbl-4}} 
\centering
\begin{tabular}{@{} c c c c c c c |l}
 $\Log\Lbol^{\mathrm{a}}$ & $\langle\Log\Lbol\rangle^{\mathrm{b}}$ & N$_{\rm obs}^{\mathrm{c}}$ & N$_{\rm corr}^{\mathrm{d}}$ & $\langle \Log\lambdaEdd \rangle$ &  $\sigma$ & $\langle \Log\lambdaEdd \rangle$\\ [0.5ex]
   [erg s$^{-1}$] & [erg s$^{-1}$] &  &  & mean &  & median \\ [0.5ex]
\hline\noalign{\smallskip}
\multicolumn{7}{c}{$z<1.2$} \\
\hline\noalign{\smallskip}
 43.00--45.03 & 44.80 & 20 & 98 & -1.47 & 0.38 & -1.66 \\[1ex]
\hline\noalign{\smallskip}
  45.03--45.48 & 45.30 & 20 & 22 & -1.15 & 0.33 & -1.18 \\[1ex]
\hline\noalign{\smallskip}
 45.48--46.50 & 45.63 & 20 & 21 & -0.78 & 0.47 & -0.83 \\[1ex]
 \hline\noalign{\smallskip}
 \multicolumn{7}{c}{$1.2\leq z\leq2.3$} \\
\hline\noalign{\smallskip}
 44.6--45.53 & 45.35 & 37 & 88 & -1.01 & 0.37 & -1.04 \\[1ex]
\hline\noalign{\smallskip}
 45.53--45.95 & 45.74 & 36 & 51 & -0.83 & 0.31 & -0.84 \\[1ex]
\hline\noalign{\smallskip}
 45.95--48.00 & 46.22 & 36 & 38 & -0.56 & 0.35 & -0.58 \\[1ex]
\hline\hline
\end{tabular}\\
\vspace{-0.2cm}\flushleft\begin{list}{}
 \item[${\mathrm{a}}$]{ Bolometric luminosity intervals.}
 \item[][${\mathrm{b}}$]{ Median Bolometric luminosity.}
 \item[][${\mathrm{c}}$]{ Number of observed objects in each bin.}
 \item[][${\mathrm{d}}$]{ Number of predicted objects in each bin employing the $V/V_{max}$ method.}
\end{list}
\end{table}
\begin{table}
\caption{Average $\Log\lambdaEdd$ values in $\Lbol-z$ bins for Type-2 AGN. \label{tbl-5}} 
\centering
\begin{tabular}{@{}c c c c c c c c |l}
 $\Log\Lbol^{\mathrm{a}}$ & $\langle\Log\Lbol\rangle^{\mathrm{b}}$ & N$_{\rm obs}^{\mathrm{c}}$ & N$_{\rm corr}^{\mathrm{d}}$ & $\langle \Log\lambdaEdd \rangle$ &  $\sigma$ & $\langle \Log\lambdaEdd \rangle$\\ [0.5ex]
   [erg s$^{-1}$] & [erg s$^{-1}$] &  &  & mean &  & median \\ [0.5ex]
\hline\noalign{\smallskip}
\multicolumn{7}{c}{$z<1.2$} \\
\hline\noalign{\smallskip}
 43.00--44.35 & 44.08 & 106 & 1317 & -1.72 & 0.59 & -1.83 \\[1ex]
\hline\noalign{\smallskip}
  44.35--44.81 & 44.62 & 106 & 234 & -1.41 & 0.50 & -1.49 \\[1ex]
\hline\noalign{\smallskip}
 44.81--46.80 & 45.08 & 105 & 140 & -0.97 & 0.57 & -1.08 \\[1ex]
 \hline\noalign{\smallskip}
 \multicolumn{7}{c}{$1.2\leq z\leq2.3$} \\
\hline\noalign{\smallskip}
 44.30--45.10 & 44.88 & 44 & 259 & -1.25 & 0.55 & -1.24 \\[1ex]
\hline\noalign{\smallskip}
 45.10--45.52 & 45.35 & 45 & 77 & -1.01 & 0.43 & -0.97 \\[1ex]
\hline\noalign{\smallskip}
 45.52--48.00 & 46.22 & 46 & 62 & -0.71 & 0.46 & -0.74 \\[1ex]
\hline\hline
\end{tabular}
\vspace{-0.2cm}\flushleft\begin{list}{}
 \item[${\mathrm{a}}$]{ Bolometric luminosity intervals.}
 \item[][${\mathrm{b}}$]{ Median Bolometric luminosity.}
 \item[][${\mathrm{c}}$]{ Number of observed objects in each bin.}
 \item[][${\mathrm{d}}$]{ Number of predicted objects in each bin employing the $V/V_{max}$ method.}
\end{list}
\end{table}

\subsection{Black hole mass-redshift dependence of Eddington ratio}
\label{Black hole mass-redshift dependence}
\begin{figure}
 \includegraphics[width=8cm]{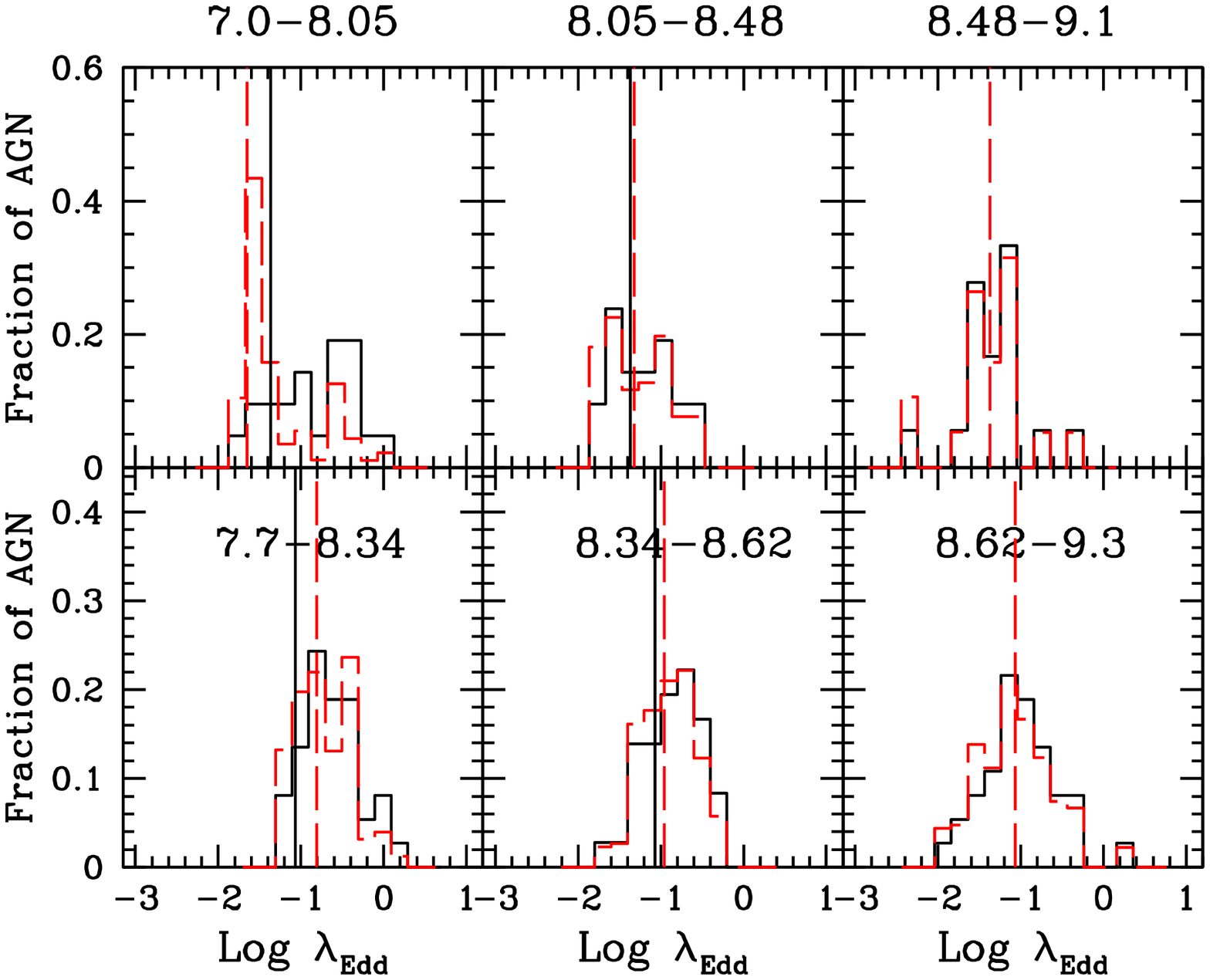}
 \caption{Distributions of Eddington ratios in bins of black hole mass and redshift for the Type-1 AGN sample. The panels are divided between $z<1.2$ (top three panels) and $1.2\leq z \leq 2.3$ (lower three panels), and black hole masses increase from left to right ($\Log \Mbh$ [$M_\odot$] intervals are reported on top of each panel). The black histograms show the observed $\lambdaEdd$ distributions, while the red dashed histograms represent the distributions corrected for incompleteness. The dashed red lines are the median values corresponding to the completeness-corrected distributions, while the solid black lines in the first and in the second bins are plotted at the $\lambdaEdd$ value corresponding to the median in the highest luminosity bin. There are $\sim20$ and $\sim37$ objects in each bin at $z<1.2$ and $1.2\leq z \leq 2.3$, respectively. The numbers of AGN in the completeness-corrected $\lambdaEdd$ distributions are 96, 26, and 19 at $z<1.2$, while there are 79, 52, and 45 at $1.2\leq z \leq 2.3$ from left to right.}
 \label{panelleddbinmbh1}
\end{figure}

\begin{figure}
 \includegraphics[width=8cm]{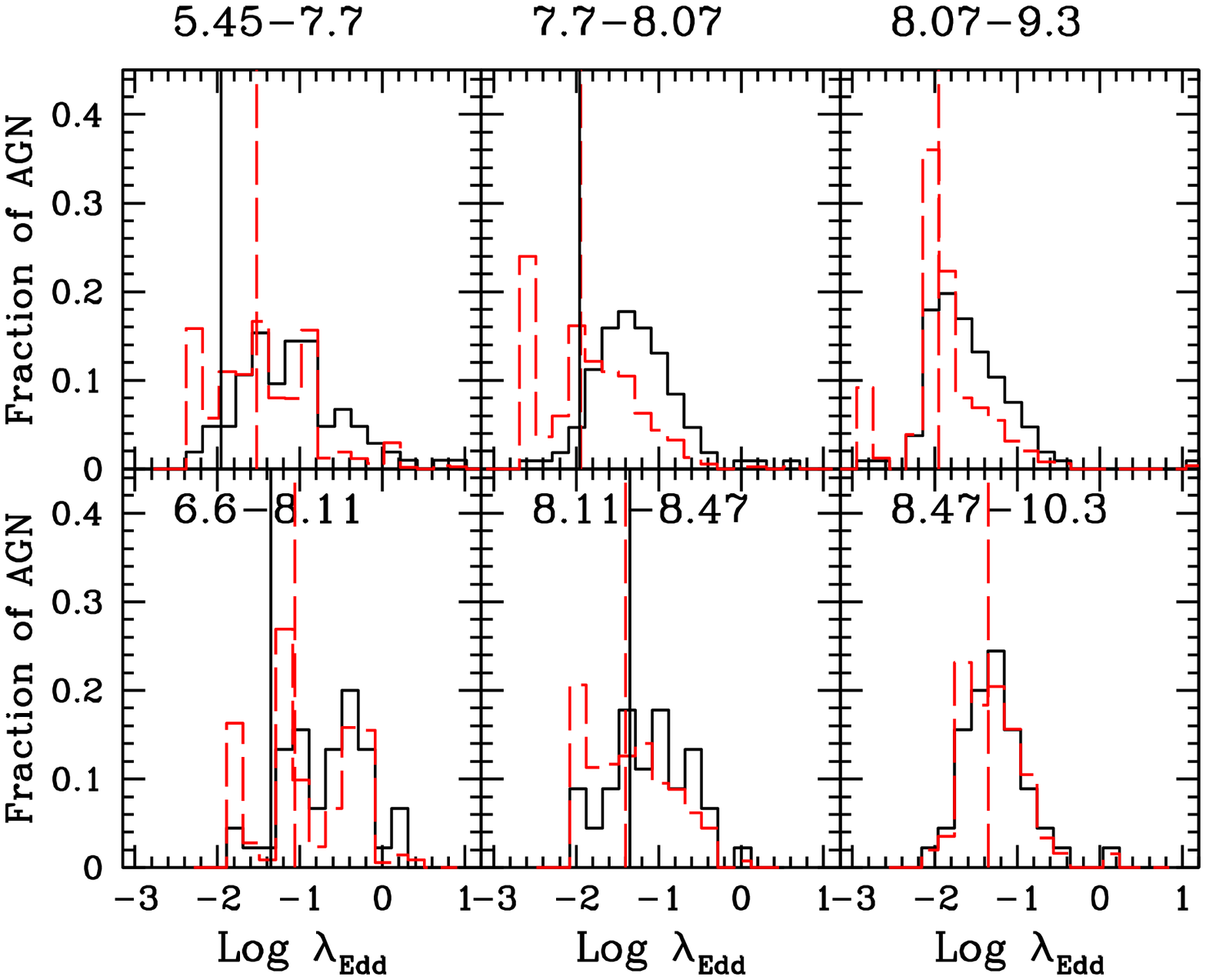}
 \caption{Distributions of Eddington ratios in bins of $\Mbh$ and redshift for the Type-2 AGN sample. Description as in Fig.~\ref{panelleddbinmbh1}. There are $\sim106$ and $\sim55$ objects in each bin at $z<1.2$ and $1.2\leq z \leq 2.3$, respectively. The numbers of AGN in the completeness-corrected $\lambdaEdd$ distributions are 1046, 384, and 262 at $z<1.2$, while there are 201, 132, and 65 at $1.2\leq z \leq 2.3$ from left to right.}
 \label{panelleddbinmbh2}
\end{figure}

The distribution of Eddington ratio as a function of black hole mass and redshift delivers more significant constraints on the physical distribution of the fueling rates.
\rev{The observed and $V/V{\rm max}$ corrected $\lambdaEdd$ distributions at a given $\Mbh$ and redshift are plotted in Figs.~\ref{panelleddbinmbh1} and \ref{panelleddbinmbh2} for the Type-1 and Type-2 AGN sample, respectively.
For both samples, each bin contains almost the same number of sources. 
From Fig.~\ref{panelleddbinmbh1} it is evident that, for Type-1 AGN, the completeness limit affects the $\lambdaEdd$ distribution only at low black hole masses in the low redshift bin. The situation for Type-2 AGN seems to be more complicated, where the low redshift bins are more affected by incompleteness in all $\Mbh$ intervals (see Fig.~\ref{panelleddbinmbh2}). 
\par
\citet{2012ApJ...746...90A} recently claimed that, for any given stellar mass, the Eddington ratio distribution of X-ray selected obscured AGN at $\Log \Lx>42$ and $z<1.0$ is well described by a power law. They reach this conclusion by running a variety of Monte Carlo simulations to correct their sample for a number of incompletenesses. Given the correlation between $M_\ast$ and $\Mbh$, the Aird et al. power-law distribution should be seen as a power-law also in bins of $\Mbh$. As shown in Figure \ref{panelleddbinmbh1} and \ref{panelleddbinmbh2}, we do not see any evidence for such a distribution in most of our redshift and $\Mbh$ bins. This is particularly clear for the Type-1 AGN (not included in Aird's analysis). The only sub-sample of Type 1 AGN, where a distribution continuously increasing toward low $\lambdaEdd$ is seen, is the sample in the low redshift and low $\Mbh$ bin. 
All the other sub-samples of Type-1 AGN show $\lambdaEdd$ distributions more consistent with Gaussians (for a similar result see also \citealt{2010MNRAS.402.2637S}). The situation is less clear-cut for the Type-2 AGN, where there are at least two bins at low redshift where the completeness-corrected distributions may suggest the presence of an underlying distribution increasing toward low $\lambdaEdd$. 
However, no clear evidence for such a power law is present in our data in the higher redshift bin for Type 2 AGN. This result is consistent with the findings of Shankar et al. (2011) where it is shown that the $\lambdaEdd$ distribution at high redshift has to be Gaussian in order to match the observed luminosity function, while at low redshift the power-law distribution is preferred.
}
\par
\begin{figure}
 \includegraphics[width=8cm]{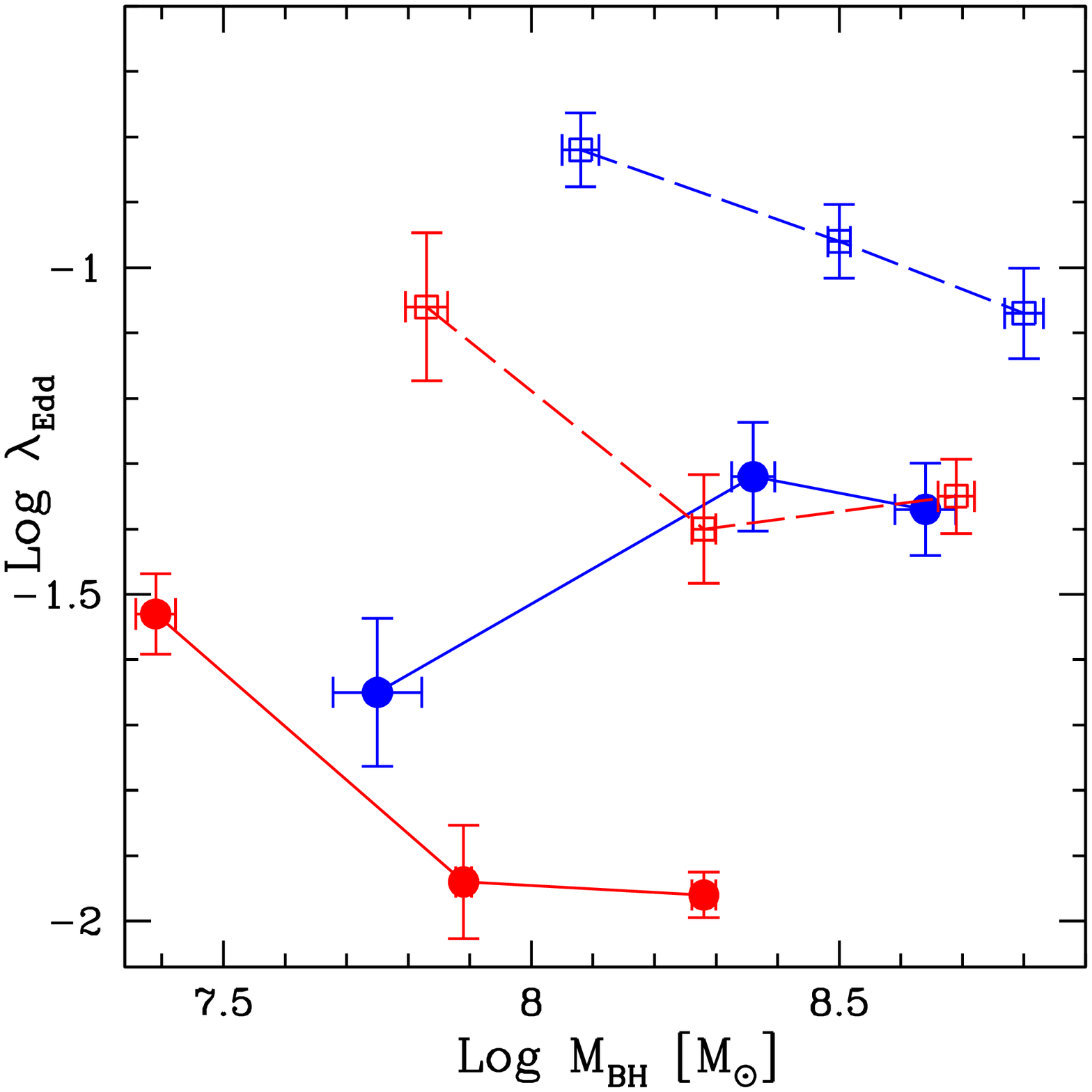}
 \caption{Median Eddington ratios as a function of the median $\Mbh$ for Type-1 (\textit{blue symbols}) and Type-2 AGN (\textit{red symbols}). Filled circles and open squares represent the median $\lambdaEdd$ for $z<1.2$ and $1.2\leq z \leq 2.3$, respectively. Solid lines connect low redshift bins, while dashed ones connect high redshift bins. Error bars on the median are estimated considering the 16th and 84th percentile divided by the square root of the observed number of AGN.}
 \label{mbhledd}
\end{figure}

The Eddington ratio as a function of $\Mbh$ is plotted in Fig.~\ref{mbhledd} for the Type-1 and Type-2 AGN samples.
The two AGN samples show higher $\lambdaEdd$ at higher redshift at any given $\Mbh$. 
In \citet{2004MNRAS.354.1020S} and \citet{2009NewAR..53...57S} it was shown that an increasing $\lambdaEdd$ with redshift may yield better results with the low mass end of the local BH mass function.
Shankar et al. (2011) also showed that an increasing $\lambdaEdd$ with redshift yields very good agreement with the high duty cycles inferred from X-ray studies at $z=0$ (e.g., \citealt{2010MNRAS.406..597G}), and with [\ion{O}{iii}] lines (e.g., \citealt{kauffmann03} and \citealt{2005MNRAS.362...25B}).
A trend of increasing $\lambdaEdd$ with redshift has also been also found by \citet{2007ApJ...654..754N} using a sample of 9818 SDSS Type 1 AGN at $z\leq0.75$. A comparison between their sample and ours is difficult given that we are sampling a different redshift range (only 18 Type-1 AGN have $z\leq0.75$ in our sample). 
However, we are in agreement with the result by Netzer and collaborators extending the analysis to higher redshifts and using a sizable X--ray selected Type-1 AGN sample. 

\section{Summary and Conclusions}
\label{Summary and Conclusions}
A homogeneous analysis of the bolometric output and Eddington ratio of 929 AGN at different X--ray absorption levels is presented. Several aspects of the present analysis have been improved with respect to L10 and L11. In particular, the far-infrared emission is now better constrained thanks to the inclusion of \textit{Herschel} data at 100~$\mu$m and 160~$\mu$m in the SED-fitting code for Type-2 AGN. Our main sample is further extended at fainter magnitudes with the addition of a sizable number of objects with photometric redshift, in order to take bias and selection effects under control. The photometric redshift catalog is the latest release provided by S11. Moreover, we have increased the coverage in the near-infrared including the H band photometry.
Black hole masses for Type-1 AGN are available for 170 sources computed from virial estimators using different lines width (\ion{Mg}{ii} and \ion{H}{$\beta$}).
Black hole masses and Eddington ratios for Type-2 AGN are estimated for 481 objects through scaling relations (\citealt{2004ApJ...604L..89H}) using a Monte Carlo method in order to account for uncertainties in $M_\ast$, $\Lbol$, as well as the intrinsic scatter in the $\Mbh-M_\ast$ relation.
We have analysed the dependence of $\kbol$ on $\Lbol$ in the B-band at 0.44~$\mu$m, in the soft and hard X--ray bands and we have compared our results with the predicted curves by M04 and H07. 
Eddington ratios are studied as a function of hard X--ray luminosities, $\Mbh$, $\Lbol$ and redshift for both Type-1 and Type-2 AGN samples taking into account incompleteness effects.
\par
Our main results are the following:
\begin{enumerate}
 \item There is a trend for higher bolometric corrections at higher bolometric luminosities in the [0.5-2]keV and [2-10]keV bands for both Type-1 and Type-2 AGN samples. The $\kbol-\Lbol$ relations by M04 and H07 predict higher $\kbol$ than what is observed in our X--ray selected AGN samples (both Type-1 and Type-2 AGN). The range of validity of these curves are limited to about two orders of magnitudes for both Type-1 ($11\leq \Log \Lbol [L_\odot]\leq 13$) and Type-2 AGN ($10\leq \Log \Lbol [L_\odot]\leq 12$).  In the overlapping luminosity range ($\Lbol=10^{11-12}L_\odot$) there is no significant difference between $\kbol$ for Type-1 and Type-2 AGN, and moreover, the $\kbol-\Lbol$ relation for Type-2 AGN seem to be the natural extension of the Type-1 relation at lower luminosities.
 \item The $\kbol-\Lbol$ relation in the B-band is in agreement with the H07 and M04 relations for $\Lbol>6\times10^{11}L_\odot$. The lowest bolometric luminosity bins do not follow the predicted trend. A fraction of our Type-1 AGN is experiencing a significant nuclear dust extinction, along with host galaxy contamination (Elvis et al 2012, submitted), and these two effects are sufficient to explain the observed difference.
 \item We confirm the trend to have higher $\kbol$ for higher Eddington ratios for both Type-1 and Type-2 AGN. The same trend has been observed with the bolometric luminosity. This indicates that the emission from the X--ray corona becomes weaker relative to the disc as the Eddington-scaled accretion rate increases.
 \item The population of AGN is dominated by sub-Eddington accretion rate objects at a given $\Lbol$.
 \item The distribution of $\Log \lambdaEdd$ is nearly Gaussian especially at high redshift and at high $\Lbol$/$\Mbh$, with a dispersion of the order of $\sim0.35$ dex for the Type-1 AGN sample and $\sim0.5$ dex for the Type-2 AGN sample. 
As expected, the low redshift/luminosity bins are more affected by incompleteness.
 \item Eddington ratio increases with bolometric luminosity for both Type-1 and Type-2 AGN.
 \item Eddington ratios show an evolution in redshift if we bin in $\Mbh$ for both AGN Types. If we instead bin in bolometric luminosity $\lambdaEdd$ do not show any clear evolution in redshift for both AGN Types.
\end{enumerate}
We want to emphasize that the $\kbol-\Lbol$ relations derived in the present work are calibrated for the first time against a sizable AGN population, and therefore rely on observed redshifts, X--ray luminosities and column density distributions. 
The application of these empirical relations offers the opportunity of future developments along several lines of investigation. For example, they could provide important hints for the computation of the black hole mass density and AGN bolometric luminosity function.
As a final comment, this analysis suggests that the fundamental physical correlation of $\kbol$ is with bolometric luminosity and Eddington ratio, rather than with single band luminosities. 

\section*{Acknowledgments}
In Italy, the XMM-COSMOS project is supported by ASI-INAF grants I/009/10/0, I/088/06 and ASI/COFIS/WP3110
I/026/07/0. In Germany the XMM-\textit{Newton} project is supported by the Bundesministerium f\"{u}r Wirtshaft und Techologie/Deutsches Zentrum f\"{u}r Luft und Raumfahrt and the Max-Planck society. M. Salvato acknowledges support by the German Deutsche Forschungsgemeinschaft, DFG Leibniz Prize (FKZ HA 1850/28-1). F. Shankar acknowledges support from a Marie Curie grant. The author acknowledges the anonymous reviewer who provided many useful suggestions for improving the paper. The entire COSMOS collaboration is gratefully acknowledged.

\bibliographystyle{mn2e}
\bibliography{bibl}

\begin{table}
\caption{Average $\Log\lambdaEdd$ values in $\Mbh-z$ bins for Type-1 AGN. \label{tbl-6}} 
\centering
\begin{tabular}{@{} c c c c c c c |l}
 $\Log\Mbh^{\mathrm{a}}$ & $\langle\Log\Mbh\rangle^{\mathrm{b}}$ & N$_{\rm obs}^{\mathrm{c}}$ & N$_{\rm corr}^{\mathrm{d}}$ & $\langle \Log\lambdaEdd \rangle$ &  $\sigma$ & $\langle \Log\lambdaEdd \rangle$\\ [0.5ex]
 [erg s$^{-1}$] & [erg s$^{-1}$] &  &  & mean &  & median \\ [0.5ex]
\hline\noalign{\smallskip}
\multicolumn{7}{c}{$z<1.2$} \\
\hline\noalign{\smallskip}
 7.00--8.05 & 7.75 & 21 & 96 & -1.31 & 0.49 & -1.65 \\[1ex]
\hline\noalign{\smallskip}
 8.05--8.48 & 8.36 & 21 & 26 & -1.28 & 0.38 & -1.32 \\[1ex]
\hline\noalign{\smallskip}
 8.48--9.10 & 8.64 & 18 & 19 & -1.38 & 0.46 & -1.37 \\[1ex]
\hline\noalign{\smallskip}
 \multicolumn{7}{c}{$1.2\leq z\leq2.3$} \\
\hline\noalign{\smallskip}
 7.70--8.34 & 8.08 & 36 & 79 & -0.73 & 0.32 & -0.82 \\[1ex]
\hline\noalign{\smallskip}
 8.34--8.62 & 8.50 & 36 & 52 & -0.90 & 0.33 & -0.96 \\[1ex]
\hline\noalign{\smallskip}
 8.62--9.30 & 8.80 & 37 & 45 & -1.06 & 0.45 & -1.07 \\[1ex]
\hline\hline
\end{tabular}
\vspace{-0.2cm}\flushleft\begin{list}{}
 \item[${\mathrm{a}}$]{ Black hole mass intervals.}
 \item[][${\mathrm{b}}$]{ Median black hole mass.}
 \item[][${\mathrm{c}}$]{ Number of observed objects in each bin.}
 \item[][${\mathrm{d}}$]{ Number of predicted objects in each bin employing the $V/V_{max}$ method.}
\end{list}
\end{table}
\begin{table}
\caption{Mean $\Log\lambdaEdd$ values in $\Mbh-z$ bins for Type-2 AGN. \label{tbl-7}} 
\centering
\begin{tabular}{@{}c c c c c c c c |l}
 $\Log\Mbh^{\mathrm{a}}$ & $\langle\Log\Mbh\rangle^{\mathrm{b}}$ & N$_{\rm obs}^{\mathrm{c}}$ & N$_{\rm corr}^{\mathrm{d}}$ & $\langle \Log\lambdaEdd \rangle$ &  $\sigma$ & $\langle \Log\lambdaEdd \rangle$\\ [0.5ex]
 [erg s$^{-1}$] & [erg s$^{-1}$] &  &  & mean &  & median \\ [0.5ex]
\hline\noalign{\smallskip}
\multicolumn{7}{c}{$z<1.2$} \\
\hline\noalign{\smallskip}
 5.45--7.70 & 7.39 & 104 & 1046 & -1.46 & 0.60 & -1.53 \\[1ex]
\hline\noalign{\smallskip}
 7.70--8.07 & 7.89 & 107 & 384 & -1.85 & 0.60 & -1.94 \\[1ex]
\hline\noalign{\smallskip}
 8.07--9.300 & 8.28 & 106 & 262 & -1.89 & 0.51 & -1.96 \\[1ex]
\hline\noalign{\smallskip}
 \multicolumn{7}{c}{$1.2\leq z\leq2.3$} \\
\hline\noalign{\smallskip}
 6.60--8.11 & 7.83 & 45 & 201 & -0.91 & 0.56 & -1.06 \\[1ex]
\hline\noalign{\smallskip}
 8.11--8.47 & 8.28 & 45 & 132 & -1.34 & 0.48 & -1.40 \\[1ex]
\hline\noalign{\smallskip}
 8.47--10.3 & 8.69 & 45 & 65 & -1.30 & 0.39 & -1.35 \\[1ex]
\hline\hline
\end{tabular}
\vspace{-0.2cm}\flushleft\begin{list}{}
 \item[${\mathrm{a}}$]{ Black hole mass intervals.}
 \item[][${\mathrm{b}}$]{ Median black hole mass.}
 \item[][${\mathrm{c}}$]{ Number of observed objects in each bin.}
 \item[][${\mathrm{d}}$]{ Number of predicted objects in each bin employing the $V/V_{max}$ method.}
\end{list}
\end{table}

\end{document}